\documentclass[prb,twocolumn,notitlepage,superscriptaddress,showkeys,showpacs]{revtex4-1}
\usepackage{amsmath,amsfonts,amssymb,amsthm,epsfig,array}
\usepackage{dsfont}
\usepackage{graphics}
\usepackage{float}
\usepackage{verbatim}
\usepackage[dvipsnames]{xcolor}
\usepackage[hidelinks,colorlinks,linkcolor=blue,
citecolor=blue,urlcolor=blue]{hyperref}
\usepackage[titletoc,title]{appendix}
\usepackage{slashed}

\newcommand{\bsl}[1]{\boldsymbol{#1}}
\newcommand{\shpa}{\shortparallel}

\newcommand{\ii}{\mathrm{i}}

\newcommand{\Tr}{\mathop{\mathrm{Tr}}}

\newcommand{\eqnref}[1]{Eq.\,\eqref{#1}}
\newcommand{\figref}[1]{Fig.\,\ref{#1}}

\newcommand{\secref}[1]{Sec.\,\ref{#1}}
\newcommand{\appref}[1]{Appendix.\,\ref{#1}}

\newcommand{\eq}[1]{\begin{equation} #1 \end{equation}}
\newcommand{\eqn}[1]{\begin{eqnarray} #1 \end{eqnarray}}
\let\oldAA\AA
\renewcommand{\AA}{\text{\normalfont\oldAA}}

\newcommand{\ie}{{\emph{i.e.} }}
\newcommand{\refcite}[1]{Ref.\,[\onlinecite{#1}]}
\newcommand{\eg}{{\emph{e.g.} }}

\begin{document}
\title{Finite-Scale Emergence of 2+1D Supersymmetry at First-Order Quantum Phase Transition}
\author{Jiabin Yu}
\email{jky5062@psu.edu}
\affiliation{Department of Physics, the Pennsylvania State University, University Park, PA, 16802}
\author{Radu Roiban}
\affiliation{Department of Physics, the Pennsylvania State University, University Park, PA, 16802}
\author{Shao-Kai Jian}
\affiliation{Institute for Advanced Study, Tsinghua University, Beijing 100084, China}
\author{Chao-Xing Liu}
\email{cxl56@psu.edu}
\affiliation{Department of Physics, the Pennsylvania State University, University Park, PA, 16802}
\begin{abstract}
Supersymmetry, a symmetry between fermions and bosons, provides a promising extension of the standard model but is still lack of experimental evidence.
Recently, the interest in supersymmetry arises in the condensed matter community owing to its potential emergence at the continuous quantum phase transition.
In this work, we demonstrate that 2+1D supersymmetry, relating massive Majorana and Ising fields, might emerge at the first-order quantum phase transition of the Ising magnetization by tuning a single parameter.
Although the emergence of the SUSY is only allowed in a finite range of scales due to the existence of relevant  masses, the scale range can be large when the masses before scaling are small.
We show that the emergence of supersymmetry is accompanied by a topological phase transition for the Majorana field,
where its non-zero mass changes the sign but keeps the magnitude.
An experimental realization of this scenario is proposed
using the surface state of a 3+1D time-reversal invariant topological superconductor with surface magnetic doping.
\end{abstract}
\maketitle

\section{Introduction}
Originally proposed
as a means to evade the Coleman-Mandula theorem and unify internal and space-time symmetries,
supersymmetry (SUSY) has been an active area of research due to its potential in
solving the hierarchy and cosmological constant problems and other puzzles in high-energy physics~\cite{Gervais1971SUSY,Wess1974SUSY,Zumino1975SUSY,Dimopoulos1981SUSY,Wess1992SUSY}.
Bosons and fermions related by SUSY transformations, referred to as superpartners, have the same mass~\cite{Wess1992SUSY};
SUSY breaking lifts this degeneracy proportionally to the breaking scale.
Despite extensive searches at high energies, conclusive experimental evidence for SUSY is yet to be found.

The last 30 years have witnessed active SUSY-related research in solid-state systems, including the tricritical Ising
model~\cite{Friedan1984SUSYTRIISing}, the boundary of topological insulators and superconductors~\cite{Hasan2010TI,Qi2010TITSC,Ponte2014SUSY,
Grover2014SUSY,Zerf2016SUSY,Witczak2016SUSY, Li2017SUSYQMC,Li2018SUSY,Jian2017SUSY}, the bulk of semimetals~\cite{Lee2007SUSY,Jian2015SUSY}, high-$T_c$ superconductors~\cite{Balents1998highTcSUSY}, the Josephson-junction array~\cite{Foda1988SUSY} and various other model systems~\cite{Thomas2005SUSY, Fendley2003SUSY,Huijse2008SUSY,Bauer2013SUSY, Rahmani2015SUSY,Huijse2015SUSY}, as well as in the cold atom system~\cite{Yu2010SUSY}.
It has been argued that, even though the corresponding microscopic models do not exhibit it,
SUSY emerges macroscopically at continuous quantum phase transitions of such solid-state systems.
The gapless nature of the continuous phase transition implies that the resulting superpartners are massless.

In this work, we present an example of emergent SUSY at the first-order quantum phase transition~(FOQPT) of a solid-state system.
The FOQPT can be achieved by tuning only one parameter and the corresponding superpartners are massive.
Although the emergent SUSY is only valid in a finite range of scales owing to the gapped nature of the FOQPT, the scale range can be large if the inital mass (mass before scaling) is small.
Specifically, we consider a 2+1D Majorana field coupled to an Ising field, and perform the one-loop renormalization group (RG) analysis in three different schemes.
Within the finite range of scales allowed by the gapped theory, all three schemes show that the FOQPTs can be reached by tuning one parameter, and have emergent SUSY with gapped Majorana and Ising fields serving as the massive superpartners.
Interestingly, the emergent SUSY is always accompanied by a topological phase transition of the Majorana field even though its mass does not vanish.
Finally, we propose an experimental realization of the emergent SUSY based on the time-reversal~(TR) invariant topological superconductor~(TSC).

\section{SUSY in a Massive Majorana-Ising System}
\label{sec:SUSY}
We consider a 2+1D action that describes a Majorana fermion $\gamma$ interacting with an Ising field $\phi$ and
discuss its SUSY.
The 2+1D action is
\eqn{
\label{eq:S}
&& S=\int d^{d}x \left[\frac{1}{2}\gamma^T(\partial_{\tau}-i v_f\bsl{\alpha}\cdot\bsl{\nabla}+m\sigma_y)\gamma+\frac{1}{2}g\phi \gamma^T\sigma_y\gamma \nonumber\right.\\
&&
\left.+\frac{1}{2}\phi(-\partial_\tau^2-v_b^2\bsl{\nabla}^2+r)\phi+a \phi+\frac{1}{3!}b\phi^3+\frac{1}{4!}u\phi^4\right]\ ,
}
where
$x=(\tau,\bsl{x})$ with the imaginary time $\tau$, and $\bsl{\alpha}=(\sigma_z,\sigma_x)$ with the Pauli matrices $\sigma_i$.
Without loss of generality, we choose $v_f > 0$ by rotating the index of the Majorana field, $g\geq 0$ by flipping the sign
of $\phi$, and $u>0$ to make the bosonic potential bounded below.
\eqnref{eq:S} has rotational invariance along $z$ and, for a uniform vacuum expectation value (VEV) of $\phi$, it is the most
general rotationally invariant action to $\phi\gamma^T\sigma_y\gamma$ and $\phi^4$ order. (See more details in \appref{app:Rota_Inv})
The action \eqref{eq:S} is not invariant under the TR transformations, $\gamma_{\tau,\bsl{x}}^T\rightarrow\gamma_{\tau,\bsl{x}}^T (i\sigma_y)$ and $\phi_{\tau,\bsl{x}}\rightarrow-\phi_{\tau,\bsl{x}}$,
unless $m, b, a=0$.
The TR-invariant case was analyzed in \refcite{Grover2014SUSY}, where it was shown that
SUSY with massless superpartners emerges by tuning the parameter $r$ to the continuous phase transition point $r=0$.

\eqnref{eq:S} has SUSY when
\eqn{
\label{eq:SUSY_rel}
&&v_f=v_b=1,\ b=3 m g,\ u=3 g^2,\nonumber\\ && a=\frac{m}{g}(r-m^2),\ r<\frac{3 m^2}{2}\ .
}
To see this, let us perform the vacuum shift $\phi=\bar{\phi}+\phi_0$ to \eqnref{eq:S} with $\phi_0$ satisfying $ a+r \phi_0+\frac{b\phi_0^2}{2}+\frac{u \phi_0^3}{3!}=0$, resulting in
\begin{eqnarray}
&& S=\int d^{d}x \left[\frac{1}{2}\gamma^T(\partial_{\tau}-i v_f\bsl{\alpha}\cdot\bsl{\nabla}+m'\sigma_y)\gamma+\frac{1}{2}g\bar{\phi} \gamma^T\sigma_y\gamma \nonumber\right.\\
&&\qquad\quad
\left.  +\frac{1}{2}\bar{\phi}(-\partial_\tau^2-v_b^2\bsl{\nabla}^2+r')\bar{\phi}+\frac{1}{3!}b'\bar{\phi}^3+\frac{1}{4!}u\bar{\phi}^4\right] \ ,
\label{eq:S_vshift}
\end{eqnarray}
where $m'=m+g\phi_0$, $r'=r+b\phi_0+\frac{1}{2}u\phi_0^2$, and $b'=b+u\phi_0$.
When the SUSY condition \eqref{eq:SUSY_rel} holds, $\phi_0$ has three different choices: $\phi_0= (-m\pm \sqrt{3 m^2-2r})/g, -m/g$, while only the first two are vacua of the \eqnref{eq:S}.
Around either of the two vacua, the SUSY condition \eqref{eq:SUSY_rel} further leads to $r'=m'^2$ and $b'=3 m' g$ in \eqnref{eq:S_vshift}, making \eqnref{eq:S_vshift} the ``real'' version of the 2+1D Wess-Zumino SUSY model.
Indeed, it is invaraint under infinitesimal SUSY transformation:
$\delta_{\xi}\bar{\phi}=\xi^T \sigma_y \gamma$ and $\delta_{\xi}\gamma=\sigma_y \alpha^\mu \xi  (i\partial_\mu \bar{\phi})+\xi (-m'\bar{\phi}-g\bar{\phi}^2/2)$, where $\xi=(\xi_1,\xi_2)^T$ are the constant Grassmann-valued parameters of the SUSY transformation~\cite{Wess1992SUSY}, $\mu=0,1,2$, $\partial_\mu=(i\partial_\tau,\bsl{\nabla})$ and $\alpha^\mu=(-\mathds{1},-\bsl{\alpha})$. (See \appref{app:SUSY} for more details.)
Therefore, \eqnref{eq:S} with \eqnref{eq:SUSY_rel} has SUSY with $\gamma$ and $\phi$ serves as superpartners.

Now the question becomes how to realize the SUSY in \eqnref{eq:S} since \eqnref{eq:SUSY_rel} is not typically satisfied.
Naively, one may think 4 parameters need to be fine-tuned to realize \eqnref{eq:SUSY_rel}, \ie tuning $(\frac{v_f}{v_b}, \frac{b}{m g}, \frac{u}{g^2}, \frac{ag}{m^3}-\frac{r}{m^2})$ to $(1,3,3,-1)$, as the velocities $v_f, v_b$ can always be chosen as 1 by rescaling the spacial coordinate once they are equal and $r<3 m^2/2$ is a parameter region instead of a critical condition.
In the following, we show through one-loop RG analysis that the first three of the above parameters can naturally flow to the SUSY-required values as the scale increases, resulting in the emergence of SUSY achievable by finely tuning only one parameter $(\frac{a g}{ m^3}-\frac{r}{m^2})$.

\section{One-loop RG Equations and Emergent SUSY at Finite Scales}

The one-loop RG analysis is performed in three schemes in $d=4-\epsilon$ dimensionss~\cite{Lee2007SUSY,Fei2016EMSUSY}: (i) dimensional regularization (DR) for \eqnref{eq:S}, (ii) DR from the so-called ``massive Gross-Neveu-Yukawa (GNY) model"~\cite{Fei2016EMSUSY}, and (iii) Wilson RG scheme with spacial momentum cutoff~\cite{Shankar1994RG} for \eqnref{eq:S}.
All three schemes show that the SUSY might emerge as the scale increases by tuning only one parameter.
Furthermore, we discuss the finite-scale nature of the emergent SUSY.

\subsection{DR and Emergent SUSY at finite Scales}

In this part, we apply DR to \eqnref{eq:S} and discuss the finite-scale nature of the emergent SUSY.
The RG equations decouple in sectors which can be studied sequentially, starting with the completely decoupled one of the bosonic and fermionic velocities:
\begin{eqnarray}
\label{eq:RG_v}
\frac{d v_b}{d l}=\frac{\widetilde{g}^2 \left(v_{f}^2-v_{b}^2\right)}{32 \pi ^2 v_{b} v_{f}^3}\ , \frac{d v_f}{d l}
=\frac{\widetilde{g}^2 (v_{b}-v_{f})}{6 \pi ^2 v_{b} (v_{b}+v_{f})^2} \ ,
\end{eqnarray}
where $l$ parametrizes the scaling $(\tau,\bsl{x})\rightarrow e^l(\tau,\bsl{x})$, $\widetilde{g}=g \widetilde{\mu}^{-\epsilon/2}$ with $\widetilde{\mu}$ independent of $l$ and having the energy unit, see {\it e.g.}~\refcite{Srednicki2007QFT}.
Structure of perturbation theory implies that  the
dimensionful parameters $m$, $b$, $r$ and $a$ cannot appear in \eqnref{eq:RG_v} in DR; thus the velocities flow stably toward $v_f=v_b\equiv v$ for any non-zero $g$, as in the TR invariant case with $r=0$.~\cite{Grover2014SUSY}
By rescaling the spacial coordinate as $\bsl{x}\rightarrow v \bsl{x} $, we can choose $v_f=v_b=1$ to study the RG flows of other parameters.\cite{Nielsen1978Lorentz,Kane1995FQHE}

\begin{figure}[t]
\includegraphics[width=\columnwidth]{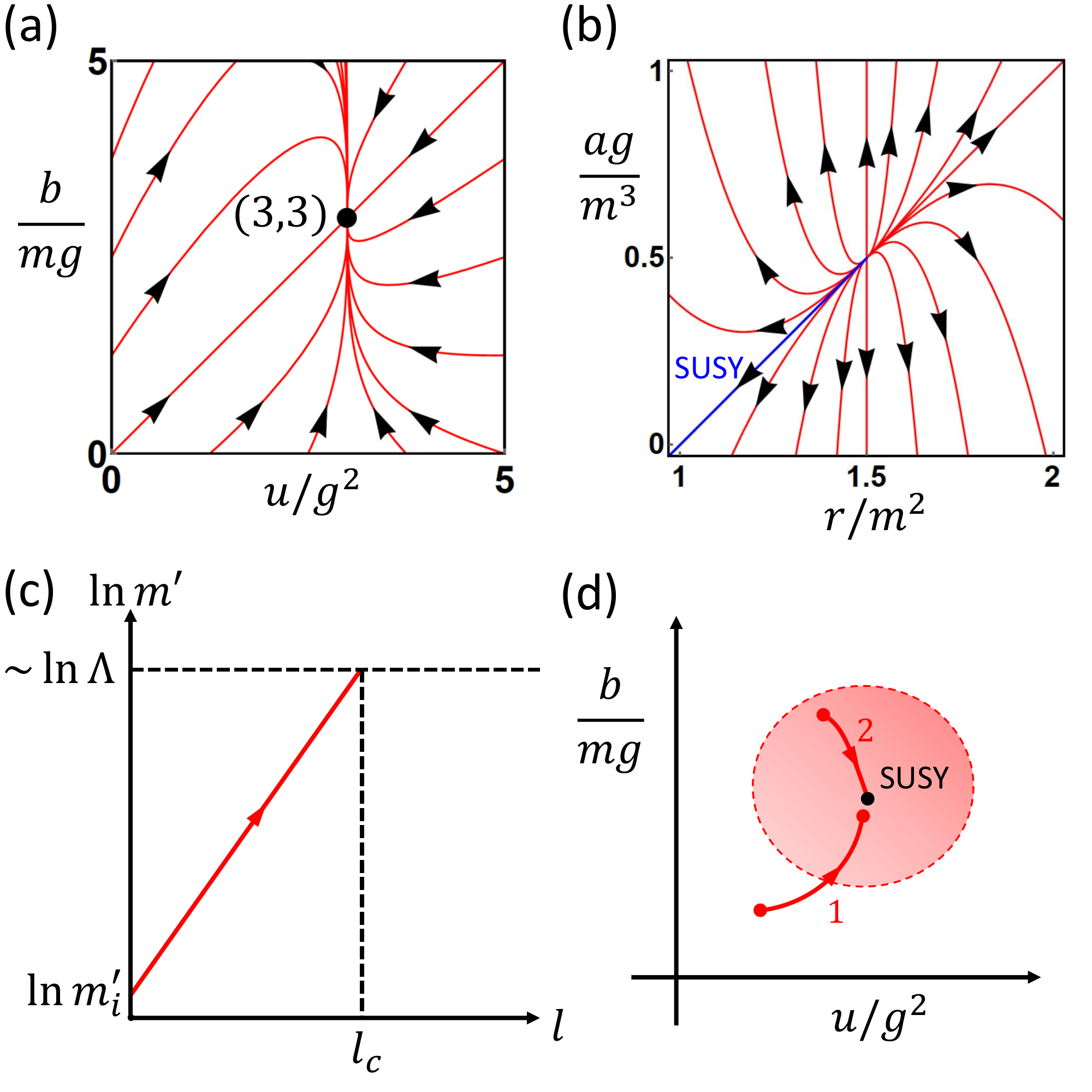}
\caption{\label{fig:RG_scheme}
(a) shows the RG flow of $b/(m g)$ and $u/g^2$ in \eqnref{eq:RG_gu} and \eqnref{eq:RG_mb} with $\widetilde{g}=\sqrt{16 \pi^2/7}$ and the black dot at $(3,3)$.
(b) is the RG flow of $a g/m^3$ and $r/m^2$ according to \eqnref{eq:ra_fixed_ub} with $\widetilde{g}=\sqrt{16 \pi^2/7}$, in which the blue line corresponds to $a g/m^3-r/m^2+1=0$ with $r<3 m^2/2$ and has SUSY.
(c) schematically shows the growth of the mass $m'$ in \eqnref{eq:S_vshift} with SUSY, where $m_i'$ is the initial mass (at $l=0$). $l_c$ indicates the critical scale at which the mass is comparable with the ultraviolet cutoff $\Lambda$. The arrow points in the $l$-increasing direction.
(d) schematically shows the RG flow to the SUSY point before $l$ reaching $l_c$. The arrow points in the $l$-increasing direction, and the red dots at tail and head correspond to $l=0$ and $l=l_c$, respectively. The black dot is on the SUSY hypersurface, and the head red dot of path 2 is missing as it is too close to the SUSY black dot. The red region indicates the SUSY-emergent region; if a system is initially in this region, the SUSY signature of the system at $l=l_c$ might be identified as exact within the numerical and experimental errors. Examples of paths 1 and 2 are shown in \figref{fig:SUSY_reg} of \appref{app:RG}4.
}
\end{figure}

We next consider the RG equations of $g$ and $u$,
\eqn{
\label{eq:RG_gu}
&& \frac{d \widetilde{g}}{d l}=\widetilde{g} \left(\frac{\epsilon }{2}-\frac{7 \widetilde{g}^2}{32 \pi ^2}\right)\nonumber\\
&&\frac{d \widetilde{u}}{d l}=\frac{12 \widetilde{g}^4-2 \widetilde{g}^2 \widetilde{u}-3 \widetilde{u}^2}{16 \pi ^2}+\widetilde{u} \epsilon\ ,
}
where $\widetilde{u}= u\widetilde{\mu}^{-\epsilon}$.
Similar to the velocities, the RG equations of $g$ and $u$ are the same as the TR invariant case at $r=0$, since they are
dimensionless in the absence of the dimensional regulator \cite{Grover2014SUSY}.
Thereby, $g$ stably flows toward a non-zero value $g^*=\sqrt{16 \pi^2 \epsilon \widetilde{\mu}^\epsilon/7}$, while $u$ stably flows toward $u^*=3 (g^*)^2$.
The RG equations for $m$ and $\widetilde{b}= b\widetilde{\mu}^{-\epsilon/2}$ are
\begin{eqnarray}
\label{eq:RG_mb}
&& \frac{d m}{d l}=m-\frac{3 \widetilde{g}^2 m}{16 \pi ^2}\nonumber\\
&&\frac{d \widetilde{b}}{d l}=\widetilde{b}(1+\frac{\epsilon}{2}) +\frac{\widetilde{b} \left(-3 \widetilde{g}^2-6 \widetilde{u}\right)+24 \widetilde{g}^3 m}{32 \pi ^2}\ .
\end{eqnarray}
For $g=g^*$, the mass anomalous dimension is $3\epsilon/7<1$.
Thus, $m$ is relevant and flows toward a value $m^{*}$ determined by the system scale.
In contrast to the conventional Ising model without TR-breaking term~\cite{Goldenfeld2018PT}, the $\phi^3$ term is
included in \eqnref{eq:S} since it can be generated by the TR-breaking $m$ term as shown below.
For a non-zero $m$, it is useful to consider the flow of $b/(m g)$ at the fixed point of the flow of $u$, \ie $u=3 g^2$.
It is given by
\begin{equation}
\frac{d }{d l}\left(\frac{b}{ m g}\right)=-\frac{\widetilde{g}^2}{4 \pi ^2}(\frac{b}{m g}-3)\ .
\end{equation}
This indicates that $b$ flows stably toward $b^{*}=3 m^{*} g^*$, and thus $b$ can be driven away from zero by a non-zero $m$ as long as $g\neq 0$.
The RG flow of $(u,b)$, stably flowing toward $(u^*,b^{*})=(3 (g^*)^2, 3 m^{*} g^*)$, is also verified by numerically
plotting the RG flows of $u/g^2$ and $b/(g m)$ in \figref{fig:RG_scheme}(a).

The RG equations of $r$ and $a$ read
\eqn{
\label{eq:RG_ra}
&&\frac{d r}{d l}=2 r-\frac{\widetilde{b}^2+\widetilde{g}^2 \left(r-6 m^2\right)+r \widetilde{u}}{16 \pi ^2}\nonumber\\
&&\frac{d \widetilde{a}}{d l}=(3-\frac{\epsilon}{2})\widetilde{a}-\frac{\widetilde{a} \widetilde{g}^2+2 \widetilde{b} r-4 \widetilde{g} m^3}{32 \pi ^2}\ ,
}
where $\widetilde{a}=a\widetilde{\mu}^{\epsilon/2}$.
Though $r$ and $a$ do not have any stable flow as suggested by the above equations, they have unstable fixed points at $r/m^2=3/2$ and $a g/m^3-r/m^2+1=0$, shown by the following RG equations with  $(u/g^2,b/(m g))=(3,3)$:
\eqn{
\label{eq:ra_fixed_ub}
&&\frac{d }{dl}\left(\frac{r}{m^2}\right)=\frac{ \widetilde{g}^2}{8 \pi^2}\left[\left(\frac{r}{m^2}\right)-\frac{3}{2}\right]\nonumber\\
&&\frac{d }{d l}\left(\frac{a g}{m^3}-\frac{r}{m^2}+1\right)=\frac{5 \widetilde{g}^2 }{16 \pi ^2}
\left(\frac{a g}{m^3}-\frac{r}{m^2}+1\right)\ ,
}
as well as \figref{fig:RG_scheme}(b).
In summary, one-loop RG analysis shows that as the scale $l$ increases, the action \eqref{eq:S} flows toward
\begin{eqnarray}
\label{eq:S*}
&& S^*=\int d^{d}x \left[
\frac{1}{2}\gamma^T(i\partial_{\mu}\alpha^{\mu}+m^{*}\sigma_y)\gamma+\frac{1}{2}g^*\phi \gamma^T\sigma_y\gamma\right.\nonumber\\
&&
\left. +\frac{1}{2}\phi(-\partial^2+r^{*})\phi+a^*\phi+\frac{1}{2}g^* m^{*}\phi^3+\frac{1}{8}(g^*)^2\phi^4\right]
\ ,
\end{eqnarray}
where the summation over repeated index is implied, $r^{*}$ and $a^{*}$ are the macroscopic values of $r$ and $a$, and $\partial^2=\partial_\tau^2+\bsl{\nabla}^2$.
Compared with \eqnref{eq:SUSY_rel}, only $(\frac{a^* g^*}{ (m^{*})^3}-\frac{r^{*}}{(m^{*})^2})$ needs to be finely tuned to $-1$ in order to achieve SUSY, and $r^{*}<3 (m^{*})^2/2$ is guaranteed if $r<3 (m)^2/2$ holds at $l=0$ as suggested by \eqnref{eq:ra_fixed_ub}, indicating that the blue line in \figref{fig:RG_scheme}(b) has SUSY.
Therefore, the SUSY can emerge as the scale increases after tuning one parameter.
(See more details on RG equations in \appref{app:RG}.)

However, unlike quantum critical points, the emergent SUSY of \eqnref{eq:S} is only valid in a finite range of scales due to the existence of relevant parameters and ultraviolet cutoff $\Lambda$ of the effective actions.
Let us first focus on the SUSY-invaraint version of \eqnref{eq:S_vshift}, since it is the expansion of SUSY-invariant version of \eqnref{eq:S} around the SUSY vacua.
In $d=4-\epsilon$ dimensions, \eqnref{eq:S_vshift} has one relevant parameter $m'$ when SUSY exists, which exponentially grows as $l$ increases from $0$, as shown in \figref{fig:RG_scheme}(c).
When $l$ reaches a critical scale $l_c$, $m'$ becomes comparable with $\Lambda$; the effective theory fails as $l$ goes beyond $l_c$.
As a result, the SUSY of the theory is only meaningful when $0<l\leq l_c$.
However, in experiments and numerical simulations, the system always has finite size, or equivalently finite scale $l$, and thus it is possible to make a sample with scale smaller than $l_c$ in order to probe the SUSY signature in this case.

Now we return to \eqnref{eq:S} and consider the case where the system is initially away from the SUSY hypersurface (\eg $b/(m g)$ and $u/g^2$ are not exactly 3 at $l=0$) while keeping $\frac{a g}{ m^3}-\frac{r}{m^2}=-1$ and $r<3 m^2/2$ by parameter tuning.
The system still flows to the SUSY hypersurface as the scale $l$ increases before reaching $l_c$, as suggested by the RG equations \eqnref{eq:RG_v},\eqref{eq:RG_gu} and \eqref{eq:RG_mb}. (See \figref{fig:RG_scheme}(d).)
The validity of one-loop RG equations also requires $l<l_c$ as discussed in \appref{app:RG}.
Reflected in experiments and numerical simulations, the SUSY signature becomes better and better as the sample size increases before reaching the critical scale $l_c$.
Moreover, the SUSY signature of the sample at the size of $l_c$ becomes better if the initial deviation from the SUSY hypersurface decreases.
Therefore, we can define a region called “SUSY-emergent region” such that if a system is initially in this region, the low energy physics of the system at the scale $l=l_c$ is controlled by the SUSY fixed point suggested by the one-loop RG equations, possibly leading to ``exact" SUSY signature within the numerical and experimental errors.
(\figref{fig:RG_scheme}(d)). 
In this way, \eqnref{eq:S*} can be interpreted as the action of a system in the SUSY-emergent region at a scale $l^*$ that is close to the critical scale $l_c$.
Since $l_c$ increases as the initial mass (\eg $m'$ in \eqnref{eq:S_vshift} at $l=0$) decreases, the SUSY-emergent region expands as the initial mass decreases, and covers the whole parameter space if the initial mass approaches zero, restoring the massless limit.
In sum, despite of the existence of the relevant parameters, the signatures of the emergent SUSY can be found for a wide range of scales in a large parameter region (though not completely generic as the massless case).

\subsection{Massive GNY model}
A concern of the RG analysis done in the last part is that the validity of RG in $d=4-\epsilon$ dimensions might be undermined by the fact that the action \eqref{eq:S} does not have 3+1D correspondence since all Pauli matrices have been used for the fermion in 2+1D.~\cite{Fei2016EMSUSY}
To resolve this issue, let us first introduce the so-called ``massive GNY model".
By adding mass-related terms and breaking the Lorentz invairance of the originally massless GNY model~\cite{Fei2016EMSUSY}, we arrive at the massive GNY model that reads
\eqn{
\label{eq:S_MGNY}
&& S_{MGNY}= \int d^d x [\bar{\Psi}_j(\partial_\tau\bar{\gamma}^0+v_f \bsl{\bar{\gamma}}\cdot\bsl{\nabla}+m)\Psi_j+g\phi\bar{\Psi}_j\Psi_j\nonumber\\
&&+\frac{1}{2}\phi(-\partial_\tau^2-v_b^2\bsl{\nabla}^2+r)\phi+a\phi+\frac{1}{3!}b\phi^3+\frac{u}{4!}\phi^4 ]\ ,
}
where $j=1,...,N_f$ is summed over, $\Psi_j$ is the four-component Dirac spinor, $N_f$ is the number of Dirac fermions, and $\bar{\gamma}^\mu$'s are the $4\times 4$ matrices that satisfy $\{\bar{\gamma}^\mu,\bar{\gamma}^\nu\}=2\delta^{\mu\nu}$.
\eqnref{eq:S_MGNY} has similar form as \eqnref{eq:S} if defining $\bar{\psi}=\gamma^T\sigma_y$, $\psi=\gamma$, $\bar{\gamma}^0=\sigma_y$ and $\bsl{\bar{\gamma}}=-\ii \sigma_y\bsl{\alpha}$ for \eqnref{eq:S}.
There are two key differences between \eqnref{eq:S} and \eqnref{eq:S_MGNY}: (i) that \eqnref{eq:S} is in 2+1D while \eqnref{eq:S_MGNY} is in 3+1D, and (ii) that \eqnref{eq:S} only has one complex fermionic degree of freedom while \eqnref{eq:S_MGNY} has $N=4N_f$ ones.
The first difference does not matter since the RG equations of both actions are derived in $d=4-\epsilon$ dimensions, and the second difference can be resolved by formally limiting $N$ to $1$ in the RG equations of \eqnref{eq:S_MGNY} as proposed in \refcite{Fei2016EMSUSY}.

The RG equations of velocities in \eqnref{eq:S_MGNY} are still decoupled from others and read
\eq{
\label{eq:GNY_RG_v}
\frac{d v_b}{d l}=\frac{N \widetilde{g}^2 \left(v_{f}^2-v_{b}^2\right)}{32 \pi ^2 v_{b} v_{f}^3}\ , \frac{d v_f}{d l}
=\frac{\widetilde{g}^2 (v_{b}-v_{f})}{6 \pi ^2 v_{b} (v_{b}+v_{f})^2} \ .
}
The above equation is exactly the same as \eqnref{eq:RG_v} after taking $N\rightarrow 1$, implying the emergence of Lorentz invariance $v_f=v_b=1$.
Similar as the last part, we study the RG flows of other parameters for $v_f=v_b=1$, and thereby the RG equations of $g$, $u$, $m$, $b$, $r$ and $a$ in \eqnref{eq:S_MGNY} take the form
\begin{eqnarray}
\label{eq:GNY_RG_gumbra}
&& \frac{d \widetilde{g}}{d l}=\widetilde{g} \left(\frac{\epsilon }{2}-\frac{(6+N) \widetilde{g}^2}{32 \pi ^2}\right)\nonumber\\
&&\frac{d \widetilde{u}}{d l}=\frac{12 N \widetilde{g}^4-2 N \widetilde{g}^2 \widetilde{u}-3 \widetilde{u}^2}{16 \pi ^2}+\widetilde{u} \epsilon\nonumber\\
&& \frac{d m}{d l}=m-\frac{3 \widetilde{g}^2 m}{16 \pi ^2}\nonumber\\
&&\frac{d \widetilde{b}}{d l}=\widetilde{b}(1+\frac{\epsilon}{2}) +\frac{\widetilde{b} \left(-3 N \widetilde{g}^2-6 \widetilde{u}\right)+24 N \widetilde{g}^3 m}{32 \pi ^2}\nonumber\\
&& \frac{d r}{d l}=2 r-\frac{\widetilde{b}^2+N \widetilde{g}^2 \left(r-6 m^2\right)+r \widetilde{u}}{16 \pi ^2}\nonumber\\
&&\frac{d \widetilde{a}}{d l}=(3-\frac{\epsilon}{2})\widetilde{a}-\frac{N\widetilde{a} \widetilde{g}^2+2 \widetilde{b} r-4 N\widetilde{g} m^3}{32 \pi ^2}\ .
\end{eqnarray}
The above RG equations of $g$ and $u$ coincide with those in \refcite{Fei2016EMSUSY} at the one-loop level since they are not influenced by the relevant parameters.
Formally taking the $N=1$ limit in \eqnref{eq:GNY_RG_gumbra} renders the exact match to \eqref{eq:RG_gu}, \eqref{eq:RG_mb} and \eqref{eq:RG_ra}, verifying the RG scheme used in the last part.

\subsection{Momentum Cutoff}

In this part, we revisit the emergent SUSY with the momentum cutoff regularization, \ie integrating the high-energy modes with spacial momentum $\bsl{k}$ satisfying $\Lambda(1-dl)<|\bsl{k}|<\Lambda$.
The finite-scale nature of the emergent SUSY is also reflected in this scheme.

Since this RG scheme does not need the counter-terms, the one-loop Feynman diagrams for this RG scheme are the same as those without counter-terms in \figref{fig:one-loop-dia}.
Before deriving the RG equations according to the diagrams, let us first redefine the following quantites: $\widetilde{g}=g \Lambda^{-\epsilon/2}$, $\widetilde{u}=u \Lambda^{-\epsilon}$, $\widetilde{b}=b \Lambda^{-\epsilon/2}$ and $\widetilde{a}=a \Lambda^{\epsilon/2}$.
We assume that the relevant parameters are much smaller than $\Lambda$ (with proper power according to the dimension of the quantity), and only keep the zeroth order of $1/\Lambda$ in the RG equations.
As a result, the RG equations of the velocity reads
\eqn{
\label{eq:RG_v_Wil}
&& \frac{d v_b}{d l}=(2\pi^2 K_d)\frac{\widetilde{g}^2 \left(v_{f}^2-v_{b}^2\right)}{32 \pi ^2 v_{b} v_{f}^3}\nonumber\\
&& \frac{d v_f}{d l}
=(2\pi^2 K_d)\frac{\widetilde{g}^2 (v_{b}-v_{f})}{6 \pi ^2 v_{b} (v_{b}+v_{f})^2} \ ,
}
where $K_d=\Omega_{d-1}/(2\pi)^{d-1}$ with $\Omega_{d-1}$ the solid angle in $d-1$ dimensions.
Since the above equations are exactly the same as \eqnref{eq:RG_v} if setting $K_d=1/(2 \pi^2)$, the velocities stably flow to the Lorentz invariant condition $v_f=v_b=1$.
With this condition, the RG equations of $\tilde{g},\tilde{u},m,\tilde{b}$, $r$ and $\tilde{a}$ are
\begin{eqnarray}
\label{eq:RG_gumbra_Wil}
&& \frac{d \widetilde{g}}{d l}=\widetilde{g} \left(\frac{\epsilon }{2}-(2\pi^2 K_d)\frac{7 \widetilde{g}^2}{32 \pi ^2}\right)\\
&&\frac{d \widetilde{u}}{d l}=(2\pi^2 K_d)\frac{12 \widetilde{g}^4-2 \widetilde{g}^2 \widetilde{u}-3 \widetilde{u}^2}{16 \pi ^2}+\widetilde{u} \epsilon\nonumber\\
&& \frac{d m}{d l}=m-(2\pi^2 K_d)\frac{3 \widetilde{g}^2 m}{16 \pi ^2}\nonumber\\
&&\frac{d \widetilde{b}}{d l}=(1+\frac{\epsilon}{2})\widetilde{b} +(2\pi^2 K_d)\frac{\widetilde{b}\left(-3 \widetilde{g}^2-6 \widetilde{u}\right)+24 \widetilde{g}^3 m}{32 \pi ^2}\nonumber\\
&&\frac{d r}{d l}=2r -(2\pi^2 K_d)\frac{ \widetilde{b}^2+\widetilde{g}^2 \left(r-6 m^2\right)+r \widetilde{u}}{16\pi^2 }+\Delta_{r}\nonumber\\
&& \frac{d \widetilde{a}}{d l}=(3-\frac{\epsilon}{2})\widetilde{a}-(2\pi^2 K_d)\frac{\widetilde{a} \widetilde{g}^2+2 \widetilde{b} r-4 \widetilde{g} m^3}{32 \pi ^2 }+\Delta_a\nonumber\ ,
\end{eqnarray}
where $\Delta_{r}=\frac{K_d}{8 }(2\widetilde{u}-4\widetilde{g}^2)\Lambda^2$ and $\Delta_a=\frac{K_d}{16 } \left(4 \widetilde{b} -8 \widetilde{g}  m \right) \Lambda^2$.
There are two differences between the above equations and those abtained from DR: (i) the $(2\pi^2 K_d)$ factor, and (ii) the extra terms $\Delta_r$ and $\Delta_a$ in the RG equations of $r$ and $a$.
Since the RG equations of $g,u,b,m$ in \eqnref{eq:RG_gumbra_Wil} have the same form as \eqnref{eq:RG_gu} and \eqref{eq:RG_mb}, the stable flow to $(u/g^2,b/(m g))=(3,3)$ is independent of $r$ and $a$.
Despite the extra terms $\Delta_r$ and $\Delta_a$ in the RG equations of $r$ and $a$, the flow of $(\frac{a g}{m^3}-\frac{r}{m^2}+1)$ for $(u/g^2,b/(m g))=(3,3)$ is similar to that in \eqnref{eq:ra_fixed_ub}, which reads
\eq{
\frac{d }{d l}\left(\frac{a g}{m^3}-\frac{r}{m^2}+1\right)=\frac{5}{8} \widetilde{g}^2 K_d \left(\frac{a g}{m^3}-\frac{r}{m^2}+1\right),
}
indicating the SUSY point is still a fixed point for the RG flow and only needs to finely tune one parameter.
As discussed at the beginning of this part, the RG equations \eqref{eq:RG_v_Wil} and \eqref{eq:RG_gumbra_Wil} are obtained in the condition that the relevant parameters $r,b,m,a$ are small compared with the cutoff $\Lambda$.
This condition again reflects the finite-scale nature of the emergent SUSY at the FOQPT discussed in the last section: the emergent SUSY can only be observed before the relevant parameters become comparable with the ultraviolet cutoff as the scale increases.

\section{Topological FOQPT}
In this section, we demonstrate that SUSY-invariant action (\eqnref{eq:S} with \eqnref{eq:SUSY_rel}) features a topological FOQPT in the sense that the Majorana field undergoes
an unusual topological phase transition, which leads to experimentally testable phenomena.
We first describe it in general and then focus on the $a^*=0$ case to elaborate the phenomenon.

As discussed in \secref{sec:SUSY}, there are two vacua for \eqnref{eq:S} when SUSY exists.
The two vacua must have the same energy since both vacua have SUSY and SUSY requires the ground state energy to be zero (after removing the constant that is decoupled to the superparnters).~\cite{Wess1992SUSY} 
This indicates that the SUSY-invariant version of \eqnref{eq:S} describes a system right at the FOQPT with the bosonic vacuum expectation value (VEV) $\langle \phi\rangle$ jumping between $(-m\pm \sqrt{3 m^2-2r})/g$.
Here we neglect the possibility that a new vacuum with lower energy appears after including all orders of quantum correction~\cite{Srednicki2007QFT}.
As the fermion mass has the expression $m_f=m+g \langle \phi\rangle$, $m_f$ takes the values $\pm \sqrt{3 m^2-2r}$ across the transition.
The sign flip of the fermion mass signals a topological phase transition, as the fermion mass domain wall can trap a 1+1D chiral Majorana mode.\cite{Wang2011TFT}
More importantly, the non-zero $m_f$ across the transition indicates the fermion gap does not close, thus representing a unique topological phase transition without gap-closing owing to its first-order transition nature.
Although a similar scenario has been discussed in the literature \cite{Amaricci2015TQPTNoGC,Roy2016TQPTNoGC,Zhu2018TQPTNoGC},
our case is special because the unchanged mass amplitude $|m_f|$ across the transition is required by the emergent SUSY.
This feature is better exposed by \eqnref{eq:S_vshift}, which is equivalent to \eqnref{eq:S}.
A suitable rewrite of the bosonic potential $V(\bar{\phi})=\frac{1}{4 !}\bar{\phi}^2 u [\frac{12}{u}(r'-\frac{1}{3}\frac{b'^2}{u})+(\bar{\phi}+\frac{2 b'}{u})^2]$ shows that the FOQPT now can occur at $r'=b'^2/(3 u)$ and $b'\neq 0$, where $\langle \bar{\phi} \rangle$ changes between $0$ and $-2b'/u$, and $m_f$ jumps between $m'$ and $m'-2 b' g /u$.
Therefore, the relation $ b' g = u m'$ imposed by SUSY is essential to maintain the unchanged magnitude of $m_f$ and flip its sign across the transition.

\begin{figure}[t]
\includegraphics[width=\columnwidth]{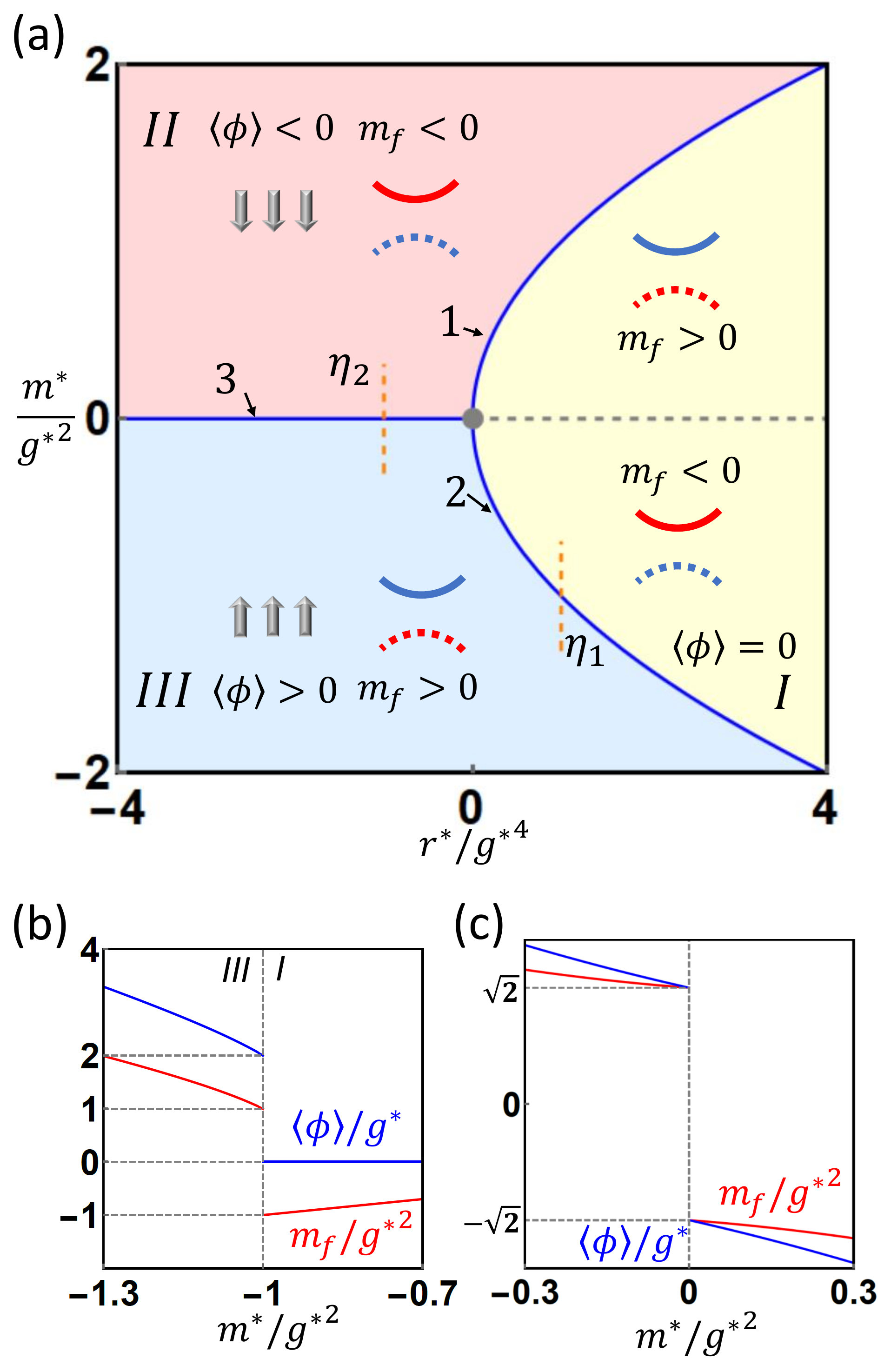}
\caption{\label{fig:PD}
(a) Phase diagram for the action \eqnref{eq:S*} with $a^*$ fixed to be zero. The blue lines 1,2 and 3 depict the topological FOQPT with emergent supersymmetry.
The gray dot at $r^*=m^*=0$ is the quantum critical point discussed in \refcite{Grover2014SUSY}.
The Ising order parameter $\langle\phi\rangle$ (silver arrows) and the dispersion of the
Majorana fermion field with the mass $m_f$ (the red and blue curves) are schematically shown for all the phases.
The Majorana field has zero mass at the gray dotted line in the phase I.
The orange dashed lines $\eta_1$ and $\eta_2$ are parameterized as $(r^*/(g^*)^4, m^*/(g^*)^2)=(1,-1+ t_1)$ and $(-1, t_1)$,
respectively, where $t_1\in [-0.3,0.3]$.
(b) and (c) show the fermion mass $m_f$ (red lines) and the VEV of the boson field $\langle  \phi\rangle$ (blue lines) when tuning the parameters along the lines $\eta_1$ and $\eta_2$ in (a), respectively.
}
\end{figure}

To better illustrate the topological FOQPT, we next discuss the action \eqnref{eq:S*} with $a^*=0$ as an example.
This case is quite general since $a^*=0$ can always be achieved by a vacuum shift like that for \eqnref{eq:S_vshift}.
We first derive the bosonic VEV $\langle \phi \rangle$ in \eqnref{eq:S*} with $a^*=0$, which stands for the macroscopic magnetic ordering, by searching for the global minimum of the bosonic part of the action.
Owing to the negative sign in front of $\partial^2$, $\langle \phi \rangle$ must be uniform in $(\tau,\bsl{x})$.
By minimizing the boson potential $V(\phi)=r^*\phi^2/2+g^* m^*\phi^3/2+(g^*)^2\phi^4/8$, we found a non-magnetic phase (I) and two magnetic
phases (II,III) with opposite values of $\langle \phi \rangle$ (See \figref{fig:PD}(a)):
\begin{eqnarray}
\label{eq:Phase}
&& \text{I}: r^*>(m^*)^2, \langle\phi\rangle=0 \\
&& \text{II}: r^*<(m^*)^2, m^*>0, \langle\phi\rangle=\frac{-3 m^* -\sqrt{9 (m^*)^2-8 r^*}}{2 g^*}\nonumber\\
&& \text{III}: r^*<(m^*)^2,m^*<0,\langle\phi\rangle=\frac{-3 m^* +\sqrt{9 (m^*)^2-8 r^*}}{2 g^*}\nonumber\ .
\end{eqnarray}
The phases I and II (III) are separated by the transition line $r^*=(m^*)^2$ with $m^*>0$ ($m^*<0$) as depicted by the line 1 (2) in \figref{fig:PD}(a),
while the Phases II and III are separated by the line $m^*=0$ and $r^*<0$ (the line 3 in \figref{fig:PD}(a)).

According to the above discussion, all the three transition lines are FOQPT lines with emergent SUSY. To show this, we consider a path across the phase transition line 1 or 2,
such as the path $\eta_1$ in \figref{fig:PD}(a). As shown by the blue line in \figref{fig:PD}(b),  $\left\langle\phi\right\rangle$ vanishes on the phase~I side, but approaches $\langle\phi\rangle=-2m^*/g^*$ as $r^*=(m^*)^2+0^-$ (on the phase II or III side).
Therefore, lines 1 and 2 are FOQPT lines with two degenerate vacua $\langle\phi\rangle=0$ and $-2m^*/g^*$, where the emergent SUSY was demonstrated above for the former vacuum.
Around the latter vacuum, the action for the boson fluctuation $\delta\phi = \phi+2m^*/g^*$ and the fermion $\gamma$ has the same form as \eqnref{eq:S*} with $a^*=0$ and the replacement $m^*\rightarrow -m^*$
and $r^*\rightarrow (m^*)^2$, which also exhibits SUSY.
Then, SUSY emerges along the FOQPT lines 1 and 2 in either of the two vacua.
Similarly, the line 3 is also a FOQPT line with two degenerate vacua $\langle \phi \rangle=\pm\sqrt{-2r^*}/g^*$ as shown by the blue line in \figref{fig:PD}(c).
Around either of the two vacua, the corresponding action for the boson fluctuation $\delta\phi=\phi-(\pm\sqrt{-2r^*}/g^*)$ and the fermion $\gamma$
can be obtained from \eqnref{eq:S*} with $a^*=0$ by replacements $m^*\rightarrow \pm \sqrt{-2 r^*}$ and $r^*\rightarrow -2 r^*$, and thus possesses SUSY.
We conclude that SUSY exists for all three FOQPT lines in \figref{fig:PD}(a) around any of the degenerate vacua.
Moreover, the topological feature of the three FOQPT lines is shown by the red lines in \figref{fig:PD}(b) and (c), where the fermion mass $m_f=m^*+g^*\langle\phi\rangle$ changes suddenly between $\pm m^*$ across the lines 1 and 2, and between $\pm\sqrt{-2r^*}$ across the line 3.

\begin{figure}[t]
\includegraphics[width=\columnwidth]{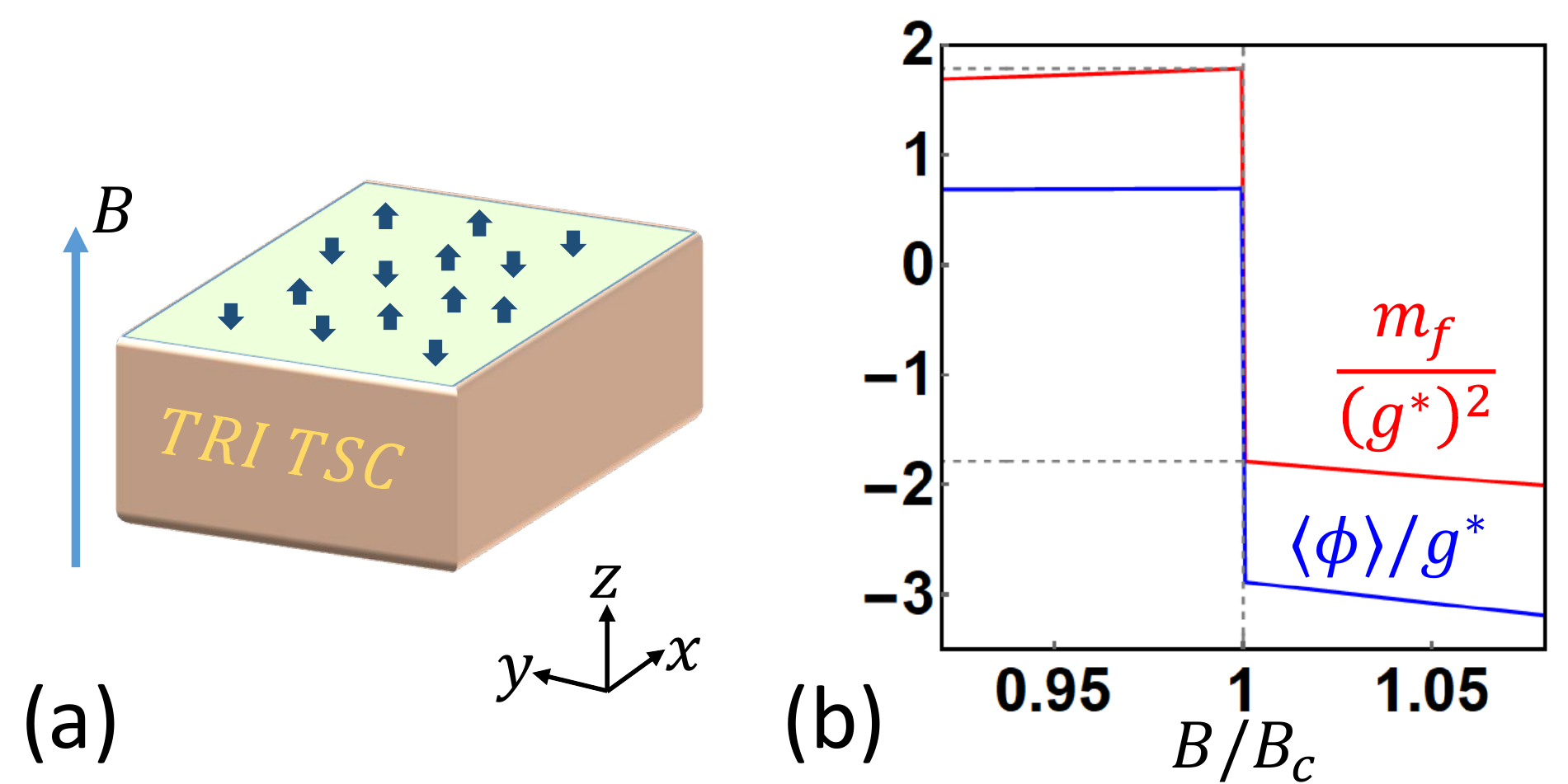}
\caption{\label{fig:setup}
(a) shows the experimental setup of TR-invariant TSC with surface magnetic doping to observe the emergent SUSY. An external magnetic field $B$ along $z$ is applied as a tuning parameter.
(b) shows the fermion mass $m_f$ (red line) and the surface magnetization $\left\langle \phi\right\rangle$ (blue line) versus the magnetic field $B$.
Here $B=B_c$ is where the FOQPT with emergent SUSY happens.
}
\end{figure}

\section{Experimental Setup}
In this section, we demonstrate that, by tuning an external magnetic field $B$, the emergent SUSY at FOQPT might be realized on the surface of a TR-invariant TSC with surface magnetic doping, as shown in \figref{fig:setup}(a).
The action of the TR-invariant TSC reads~\cite{Roy2008TRISF,Qi2009TRISCSF,Chung2009MZMHe3}
\begin{eqnarray}
\label{eq:S_3D}
&&S_0=\int \frac{d\tau d k^3}{(2\pi)^3}\left[\bar{\psi}_{\tau,\bsl{k}}(\partial_\tau +h(\bsl{k}))\psi_{\tau,\bsl{k}}+\frac{1}{2}\bar{\psi}_{\tau,\bsl{k}}\Delta_{\bsl{k}}(\bar{\psi}_{\tau,-\bsl{k}})^T\right.\nonumber\\
&&\left.+\frac{1}{2}\psi^T_{\tau,-\bsl{k}}\Delta^\dagger_{\bsl{k}} \psi_{\tau,\bsl{k}} \right]
\end{eqnarray}
where $\bar{\psi}_{\tau,\bsl{k}}=(\bar{\psi}_{\tau,\bsl{k},\uparrow},\bar{\psi}_{\tau,\bsl{k},\downarrow})$ is the Grassman field for the electron with the momentum $\bsl{k}=(k_x,k_y,k_z)$,
$h(\bsl{k})=\frac{\bsl{k}^2}{2m_0}-\mu$ with the chemical potential $\mu$, and
$\Delta_{\bsl{k}}=\Delta_p (\bsl{k}\cdot\bsl{s}) i s_y$ with the p-wave pairing $\Delta_{p}>0$ and the Pauli matrices $s_i$ for spin.
\eqnref{eq:S_3D} may be used to describe the Ce-based heavy fermion SCs and half-Heusler SCs, and its superfluid version
has been realized in B phase of He-3.~ \cite{bauer2012non,Savary2017j=3/2SC,Leggett1975He3}
For $\mu m_0 >0$, one can solve \eqnref{eq:S_3D} with an open boundary condition at $z=0$ and obtain gapless Majorana
modes $\gamma$ at the surface \cite{Roy2008TRISF,Qi2009TRISCSF,Chung2009MZMHe3}.
(See Appendix.\,\ref{app:ES} for details.)

The surface magnetic doping of TSC can be phenomenologically described by the standard Ginzburg-Landau free energy of Ising magnetism, $S_M=\int d^d x[\frac{1}{2} \phi(-\partial_\tau^2-v_b^2\bsl{\nabla}^2+r_0) \phi+\frac{1}{4!}u \phi^4]$, where $\phi_{\tau,x,y}$ is the order parameter of surface Ising magnetism along $z$.
$\phi$ is coupled to electrons at the surface through the exchange interaction, which takes the form $\frac{1}{2}g \int d^d x \phi \gamma^T \sigma_y \gamma$ after the surface projection.
Furthermore, a magnetic field $B$ along $z$ is applied and coupled to both electron spin and Ising magnetism on the surface through the Zeeman-type action, and with that, we arrive at the total action:
\begin{eqnarray}
\label{eq:S_setup}
&& S_E=\int d^{d}x \left[\frac{1}{2}\gamma^T(\partial_{\tau}-i v_f\bsl{\alpha}\cdot\bsl{\nabla})\gamma+\frac{1}{2}g \phi \gamma^T\sigma_y\gamma+\right.\nonumber\\
&&\left.\frac{1}{2}\phi(-\partial_\tau^2-v_b^2\bsl{\nabla}^2+r_0)\phi+\frac{1}{4!}u \phi^4+\frac{1}{2}\mu_B B\gamma^T\sigma_y\gamma\right.\nonumber\\
&&\left.+a_1 B \phi+\frac{a_2}{3!}B \phi^3\right] \ .
\end{eqnarray}
(See more details in \appref{app:ES}.)
Here, we neglect the orbital effect as all fields are charge neutral, and add an $\phi^3$ term since it is allowed by symmetry and can be generated at the quantum level.
Since \eqnref{eq:S_setup} has exactly the same form as \eqnref{eq:S}, its RG equations are the same as \eqnref{eq:RG_v},\eqref{eq:RG_gu}, \eqref{eq:RG_mb} and \eqref{eq:RG_ra}, resulting in $a_2^*=3 \mu_B^* g^*$ in addition to $v_f=v_b=1$ and $u^*=3(g^*)^2$.
For simplicity, we neglect the $B$-dependence of $\mu_B^*$, $r_0^*$ and $a_1^*$.
In this case, as long as $r_0^*> \frac{3 a_1^* g^*}{\mu_B^*}$, there exists a critical magnatic field $B_c=\pm \frac{1}{\mu_B^*}\sqrt{r_0^*-\frac{a_1^*g^*}{\mu_B^*}}$ such that FOQPT with emergent SUSY happens at $B=B_c$.
To demonstrate this possibility, we choose the values of parameters as $r_0^*/(g^*)^4=0.2$, $a_1^*/g^*=-1$, $\mu_B^* (g^*)^2=1$ and $B\geq 0$.
As $B$ increases to the critical value $B_c \approx 1.1 (g^*)^4$,
the FOQPT is reached.
A signature of emergent SUSY at FOQPT is that the fermion mass $m_f=m^*+g^*\langle \phi \rangle$ should have
unchanged magnitude and flip sign across the transition, verified by the red line in \figref{fig:setup}(b).
The unchanged amplitude of $|m_f|$ can be confirmed by local density of states measurement with scanning tunneling microscopy, and
the sign flip can be tested by checking the resulting topological phase transition, \ie the appearance or disappearance of chiral 1+1D domain-wall fermion.
Moreover, the surface magnetization $\langle \phi\rangle$ has a sudden change (see the blue line in \figref{fig:setup}(b))
as an evidence of FOQPT, which can be measured by superconducting quantum interference devices.

\section{Conclusion and Discussion}
In conclusion, SUSY with massive superpartners can emerge at the topological FOQPT occurring on the surface of a TR-invariant TSC in a tunable external magnetic field.
Although the emergence of the SUSY only happens in a finite range of scales owing to the gapped nature of FOQPT, the scale range can be large when the initial mass is small.
Although the emergent SUSY with massive superpartners was proposed in a cold-atom system with spontaneous symmetry breaking~\cite{Yu2010SUSY},
similar to that at line 3 in \figref{fig:PD}(a), that proposal requires to tune more than one parameter and does not discuss the relation between SUSY and the topological FOQPT.
Our work also helps shed light upon other emergent symmetries of a FOQPT.\cite{Zhao2018O4FOQFT}

\section{Acknowledgments}
J.Y. thanks Zhen Bi, Shinsei Ryu, Ashvin Vishwanath, Juven Wang and Igor Klebanov for helpful discussions. J.Y. and C.-X.L. acknowledges the support of the Office of Naval Research (Grant No. N00014-18-1-2793), the U.S. Department of Energy (Grant No.~DESC0019064) and Kaufman New Initiative research grant KA2018-98553 of the Pittsburgh Foundation. R.R.~is supported by the U.S. Department of Energy (Grant No.~DE-SC0013699). S.-K. J. is supported by the NSFC under grant 11825404.

\appendix

\section{Rotational Invariance of \eqnref{eq:S}}

\label{app:Rota_Inv}

In this section, we show that \eqnref{eq:S} of the main text is the most general rotationally invariant action to $\phi\gamma^T\sigma_y\gamma$ and $\phi^4$ order if assuming $\phi$ has uniform classical vacuum.

\begin{table}[h]
\begin{center}
\begin{tabular}{|c|c|}
\hline
$L_z$ & Expressions \\
\hline
0 & $\partial_\tau,\partial_\tau ^2,\sigma_0,\sigma_y,\bsl{\nabla}^2$ \\
\hline
-1 & $\partial_x- i \partial_y,\sigma_z-i\sigma_x$ \\
\hline
1 & $\partial_x+i\partial_y, \sigma_z+i\sigma_x$\\
\hline
-2 & $(\partial_x-i\partial_y)^2$ \\
\hline
2 & $(\partial_x+i\partial_y)^2$\\
\hline
\end{tabular}
\end{center}
\caption{\label{tab:SO2}Classification of partial derivatives and Pauli matrices according to the angular momentum $L_z$.}
\end{table}

The rotation transformation along $z$ is defined as $\gamma^T_{\tau,\bsl{x}}\rightarrow \gamma^T_{\tau,R_{\theta}\bsl{x}}e^{-i\sigma_y \theta/2}$ and $\phi_{\tau,\bsl{x}}\rightarrow \phi_{\tau,R_{\theta}\bsl{x}}$, where $R_{\theta}$ is the rotational matrix along $z$ for angle $\theta$ counterclockwise.
According to the transformation, the Pauli matrices and the derivatives in the action can be classified according to their angular momentum $L_z$, as shown in Tab.\ref{tab:SO2}.
Based on Tab.\ref{tab:SO2}, the most general action without derivatives is
\begin{equation}
S_0=\int d\tau d^{d-1} x \left[ \sum_{n\geq 0} g_n \phi^n (\gamma^T\sigma_y \gamma)+ \sum_{n\geq 1}u_n\phi^n \right]\ ,
\end{equation}
where Hermiticity requires $g_n,u_n\in \mathds{R}$, and the Hermitian conjugate transformations of the fields are $\gamma_{\tau,\bsl{x}}^{\dagger}=\gamma_{-\tau,\bsl{x}}^T$ and $\phi_{\tau,\bsl{x}}^{\dagger}=\phi_{-\tau,\bsl{x}}$ since $\tau$ is imaginary time.
In $S_0$, $u_1$ is the linear $\phi$ term, $g_0,u_2$ are the mass terms, and all other terms are the on-site interaction terms.
The most general kinetic term of $\gamma$ with leading order derivatives is
\eqn{
&& S_1=\int d\tau d^{d-1} x \gamma^T\left[ A \partial_\tau +i B_1 (\sigma_z+i\sigma_x)(\partial_x-i\partial_y)\right.\nonumber\\
&&\left.+i B_1^* (\sigma_z-i\sigma_x)(\partial_x+i\partial_y) \right]\gamma\ ,
}
where $A\in\mathds{R}$.
The most general kinetic term of $\phi$ with leading order derivatives is
\begin{equation}
S_2=\int d\tau d^{d-1} x \phi\left[ -C_0 \partial_\tau^2-C_1\bsl{\nabla}^2 \right]\phi\ ,
\end{equation}
where $C_0,C_1\in\mathds{R}$ and there are no first order derivatives since $[\phi_{\tau,\bsl{x}},\phi_{\tau',\bsl{x}'}]=0$.
Then, $S'=S_0+S_1+S_2$ is the most general rotationally invariant action if the kinetic term only contains the leading order derivatives and the on-site interaction terms have no derivatives.

Next we show how to derive \eqnref{eq:S} of the main text from $S'$.
Since we want the classical vacuum of $\phi$ to be uniform, we assume $C_0>0,C_1\geq 0$.
We also assume $A\neq 0$ which is typically true for a legitimate fermion action.
By changing integration variable $\tau\rightarrow \text{sgn}(A)\tau$ and defining $B_1= - \frac{1}{2}|A| v_f e^{i\theta_B}$, $g_n=\widetilde{g}_n |A| (C_0)^{n/2}$, $u_n=\widetilde{u}_n C_0^{n/2}$, $C_1=v_b^2 C_0$, $\gamma^T_{\text{sgn}(A)\tau,\bsl{x}}=\frac{1}{\sqrt{2|A|}}\gamma'^T_{\tau,\bsl{x}}e^{-i\sigma_y \theta_B/2}$ and $\phi_{\text{sgn}(A)\tau,\bsl{x}}=\phi'_{\tau,\bsl{x}}/\sqrt{2 C_0}$,
$S'$ becomes
\eqn{
\label{eq:S_gen}
&& S'=\int d\tau d^{d-1}x \left[ \frac{1}{2}\gamma^T(\partial_\tau-iv_f\bsl{\alpha}\cdot \bsl{\nabla})\gamma
\right.\nonumber\\
&&-\frac{1}{2}\phi(\partial_\tau^2+v_b^2\bsl{\nabla}^2)\phi+\sum_{n\geq 0}\widetilde{g}_n\frac{1}{2^{n/2+1}}\phi^n\gamma^T\sigma_y\gamma\nonumber\\
&&\left.+\sum_{n\geq 1 }\widetilde{u}_n\frac{1}{2^{n/2}}\phi^n
\right]\ ,
}
where $v_f\geq 0$ can be chosen by adjusting $\theta_B$, and $\gamma'$ and $\phi'$ are renamed as $\gamma$ and $\phi$ since they are integrated over in the partition function.
The renaming of $\gamma'$ and $\phi'$ would not cause any physical confusion since $\gamma'$ and $\phi'$ behave the same as $\gamma$ and $\phi$ under rotation, TR and hermitian conjugate:
\begin{eqnarray}
&&\gamma'^T_{\tau,\bsl{x}}=\sqrt{2|A|}\gamma^T_{\text{sgn}(A)\tau,\bsl{x}}e^{i\sigma_y \theta_B/2}
\nonumber\\
&&\overset{SO(2)}{\longrightarrow}
\sqrt{2|A|}\gamma^T_{\text{sgn}(A)\tau,R_\theta\bsl{x}}e^{-i\sigma_y \theta/2}e^{i\sigma_y \theta_B/2}\nonumber\\
&&=\gamma'^T_{\tau,R_\theta\bsl{x}}e^{-i\sigma_y \theta/2}\ ,\nonumber\\
&&\gamma'^T_{\tau,\bsl{x}}=\sqrt{2|A|}\gamma^T_{\text{sgn}(A)\tau,\bsl{x}}e^{i\sigma_y \theta_B/2}
\nonumber\\
&&\overset{TR}{\longrightarrow}
\sqrt{2|A|}\gamma^T_{\text{sgn}(A)\tau,\bsl{x}}(i\sigma_y)e^{i\sigma_y \theta_B/2}\nonumber\\
&&=\gamma'^T_{\tau,\bsl{x}}(i\sigma_y)\ ,\nonumber\\
&&(\gamma'^T_{\tau,\bsl{x}})^{\dagger}=\sqrt{2|A|}e^{-i\sigma_y \theta_B/2}\gamma_{-\text{sgn}(A)\tau,\bsl{x}}
=\gamma'_{-\tau,\bsl{x}}\nonumber\ ,
\end{eqnarray}
and obvious for $\phi'$ and $\phi$ due to their simple relation $\phi'_{\tau,\bsl{x}}=\sqrt{2 C_0}\phi_{\text{sgn}(A)\tau,\bsl{x}}$.
If we only keep the on-site interaction terms up to the $\phi\gamma^T\sigma_y\gamma$ and $\phi^4$ order  and define $\widetilde{g}_0=m$,$\widetilde{g}_1=\sqrt{2}g$, $\widetilde{u}_1=\sqrt{2}a$, $\widetilde{u}_2=r$, $\widetilde{u}_3=\sqrt{2} b / 3$ and $\widetilde{u}_4=u / 3!$,
\eqnref{eq:S_gen} is the same as \eqnref{eq:S} .
Therefore, \eqnref{eq:S} is the most general rotational invariant action if (i) only keeping leading order derivatives in the kinetic terms, (ii) neglecting derivatives in the on-site interaction, (iii) only keeping terms to $\phi\gamma^T\sigma_y\gamma$ and $\phi^4$ order for the on-site interaction, and (iv) assuming the classical vacuum of $\phi$ is uniform.

At last, we show \eqnref{eq:S} is in the most general rotationally invariant form if only keeping terms to $\phi\gamma^T\sigma_y\gamma$ and $\phi^4$ order for the on-site interaction and assuming the classical vacuum of $\phi$ is uniform, which is the statement at the beginning of this section.
It means that we need to argue why higher-order derivatives in the kinetic energy terms and derivatives in the on-site interaction can be neglected.
The argument will be done by dimension analysis.
The dimensions of fields are $[\gamma]=(d-1)/2$ and $[\phi]=(d-2)/2$.
As a result, we have $[v_f]=[v_b]=0$, $[\widetilde{u}_n]=(2-n)d/2+n$ and $[\widetilde{g}_n]=-n d/2+n+1$.
The zero dimension of $v_f$ and $v_b$ means any higher-order derivatives in the kinetic terms of $\gamma$ and $\phi$ are irrelevant, and thus can be neglected.
Now we discuss the on-site interaction.
Since we perform RG analysis in $d=4-\epsilon$, we may analyze the dimension of the interaction for $d=4$, resulting that $[\widetilde{u}_n]=4-n$ and $[\widetilde{g}_n]=1-n$.
It means, keeping $\phi\gamma^T\sigma_y\gamma$ and $\phi^4$ order is equivalent to neglect all the irrelevant on-site interaction terms at $d=4$.
The remaining on-site interaction terms include $b$, $g$ and $u$.
$g$ and $u$ are marginal, and thus adding derivatives to the two terms would make them irrelevant.
$[b]=1$ for $d=4$ and thus allows one derivative.
However, this derivative would be a total derivative since $(\partial_x\phi)\phi^2=(\partial_x\phi^3)/3$ and can be neglected.
Therefore, the derivatives in the on-site interaction can be neglected if only keeping terms to the $\phi\gamma^T\sigma_y\gamma$ and $\phi^4$ order.
The statement at the beginning of this section is proven.

\section{SUSY}
\label{app:SUSY}
In this section, we discuss the SUSY algebra and transformation.
Since the fields considered here are one two-component Majorana field and one Ising field, there should be two supercharges $Q_a$ with $a=1,2$ instead of four for $N=1$ Wess-Zumino model in 3+1D. \cite{Wess1992SUSY}
The supercharges satisfy
\begin{equation}
\left\{ Q_{a_1}, Q_{a_2}\right\}=2\bar{\alpha}^{\mu}_{a_1 a_2} P_\mu\ ,
\end{equation}
where $\bar{\alpha}^{\mu}=\sigma_y(\alpha^\mu)^T\sigma_y$ and $P_\mu$ is the energy-momentum operator that gives $[P_\mu,\varphi]=i\partial_\mu \varphi$ for any field operator $\varphi$.
Here the metric is chosen as $(-,+,+)$.
The infinitesimal SUSY transformation is defined as
\begin{equation}
\delta_\xi \varphi=-i[\xi^T\sigma_y Q,\varphi]\ ,
\end{equation}
which gives
\begin{equation}
\label{eq:SUSY_cl}
(\delta_\xi\delta_\eta-\delta_\eta\delta_\xi)\phi=2i\eta^T\alpha^\mu\xi\partial_\mu \varphi\ .
\end{equation}
Here $\xi,\eta$ are two two-component Grassmann numbers.
\eqnref{eq:SUSY_cl} shows the closure of the SUSY algebra.

Now, we show the SUSY of SUSY-invaraint version of \eqnref{eq:S_vshift}.
For simplicity, we replace $m'$ and $\bar{\phi}$ by $m$ and $\phi$, respectively.
The action can be re-written as
\begin{equation}
\label{eq:S_SUSY}
S_{SUSY}= S_{\gamma,0}+S_{\gamma\phi}+S_{\phi,0}+S_{\phi,3}+S_{\phi,4}\ ,
\end{equation}
where
\begin{eqnarray}
&& S_{\gamma,0}=\int d\tau d^{d-1}x
\frac{1}{2}\gamma^T(i\partial_{\mu}\alpha^{\mu}+m\sigma_y)\gamma\nonumber\\
&&S_{\gamma\phi}=\int d\tau d^{d-1}x \frac{1}{2}g\phi \gamma^T\sigma_y\gamma\nonumber\\
&&S_{\phi,0}=\int d\tau d^{d-1}x \frac{1}{2}\phi(-\partial^2+m^2)\phi\nonumber\\
&&S_{\phi,3}=\int d\tau d^{d-1}x \frac{1}{2}g m\phi^3\nonumber\\
&&S_{\phi,4}=\int d\tau d^{d-1}x \frac{1}{8}g^2\phi^4\ .
\end{eqnarray}
The SUSY transformation in this case can be defined as
\begin{equation}
\delta_{\xi}\phi=\xi^T \sigma_y \gamma\ ,\ \delta_{\xi}\gamma=\sigma_y \alpha^\mu (i\partial_\mu)\xi \phi+\xi (-m\phi-g\phi^2/2)\ .
\end{equation}
To demonstrate $S^*$ is SUSY invariant, we can first act $\delta_{\xi}$ on $S_{\gamma,0}$ and get
\begin{equation}
\delta_{\xi}S_{\gamma,0}=-\delta_\xi (S_{\phi,0}+\frac{1}{3}S_{\phi,3})+\int d\tau d^{d-1}x \frac{-g \phi^2}{2}\xi^T i\partial_\mu\alpha^\mu\gamma\ .
\end{equation}
Then act $\delta_{\xi}$ on $S_{\gamma\phi}$ and get
\begin{equation}
\delta_{\xi}S_{\gamma\phi}=-\delta_\xi (\frac{2}{3}S_{\phi,3}+S_{\phi,4})+\int d\tau d^{d-1}x \frac{g \phi^2}{2}\xi^T i\partial_\mu\alpha^\mu\gamma\ .
\end{equation}
As a result, we have
\begin{equation}
\delta_{\xi}(S_{\gamma,0}+S_{\gamma\phi})=-\delta_\xi (S_{\phi,0}+S_{\phi,3}+S_{\phi,4})\Leftrightarrow \delta_\xi S^*=0\ .
\end{equation}
The defined SUSY transformation must be close, i.e. satisfying \eqnref{eq:SUSY_cl}.
It is true for $\phi$, i.e.
\begin{equation}
(\delta_\xi \delta_\eta-\delta_\eta \delta_\xi)\phi=2 i \eta^T \alpha^{\mu} \xi \partial_\mu \phi\ .
\end{equation}
For $\gamma$, we have
\eqn{
&& (\delta_\xi \delta_\eta-\delta_\eta \delta_\xi)\gamma=2 i \eta^T \alpha^{\mu} \xi \partial_\mu \gamma\nonumber\\
&&-(\eta\xi^T-\xi \eta^T)(i\alpha^\mu \partial_\mu+m\sigma_y+g\phi\sigma_y)\gamma\ ,
}
where $\eta\xi^T-\xi \eta^T=\sum_{\mu} \eta^T \alpha^\mu \xi \alpha^\mu $ and $\sigma_y\alpha^\mu\alpha^\nu+\alpha^\nu\alpha^\mu\sigma_y=2\sigma_y\delta^{\mu\nu}$ are used.
Clearly, the closure of the algebra for $\gamma$ requires the equatin of motion of $\gamma$, which is $(i\alpha^\mu \partial_\mu+m\sigma_y+g\phi\sigma_y)\gamma=0$.
We call the algebra is close through the equation of motion.
The requirement of the equation of motion is because we integrate out the auxiliary field.\cite{Wess1992SUSY}

\section{Details For RG Equations}
\label{app:RG}

In this section, we derive  \eqnref{eq:RG_v},\eqref{eq:RG_gu}, \eqref{eq:RG_mb} and \eqref{eq:RG_ra}.
We first derive the Callan-Symanzik equation for $N$-point function, and then show the RG equations to the one-loop order.

\subsection{Callan-Symanzik Equation}
The regularization scheme chosen here is the dimensional regularization with $d=4-\epsilon$.
For a generic dimension $d$, the dimensions of the fields, velocities, masses and interaction couplings are $[\gamma]=(d-1)/2$, $[\phi]=(d-2)/2$, $[v_b]=[v_f]=0$, $2[m]=[r]=2$, $[a]=1+d/2$, $[g]=(4-d)/2$, $[b]=(6-d)/2$ and $[u]=4-d$.
For $d=4$, $[a]=3$, $[g]=[u]=0$ and $[b]=1$.
In order to keep the dimensions of the couplings the same as $d=4$ in the $d=4-\epsilon$ scheme, we introduce a parameter $\widetilde{\mu}$ with $[\widetilde{\mu}]=1$ by doing the transformation
\begin{equation}
\label{eq:gub2mu}
a\rightarrow a\widetilde{\mu}^{-\epsilon/2},\ g\rightarrow g\widetilde{\mu}^{\epsilon/2},\ b\rightarrow b\widetilde{\mu}^{\epsilon/2}\ \text{and}\ u\rightarrow u\widetilde{\mu}^{\epsilon}\ .
\end{equation}
In addition, in order to include the quantum corrections, we should introduce $Z$ factors to the action, and the action becomes
\begin{eqnarray}
\label{eq:S_Z}
&&S=\int d^d X \left\{
\frac{1}{2}\gamma^T[Z_\gamma \partial_{\tau}+Z_{v_f} v_f(-i\bsl{\alpha}\cdot\bsl{\nabla})+Z_m m\sigma_y]\gamma\right.\nonumber\\
&&\left.+\frac{1}{2}Z_g g\widetilde{\mu}^{\epsilon/2}\phi \gamma^T\sigma_y\gamma+\frac{1}{2}\phi(-Z_\phi \partial_\tau^2-Z_{v_b} v_b^2\bsl{\nabla}^2+Z_r r)\phi\right.\nonumber\\
&&\left.+Z_a a \widetilde{\mu}^{-\epsilon/2} \phi+\frac{1}{3!}Z_b b\widetilde{\mu}^{\epsilon/2}\phi^3+\frac{1}{4!}Z_u u\widetilde{\mu}^{\epsilon}\phi^4\right\}
\ ,
\end{eqnarray}
where the partition function is $Z=\int D\gamma D\phi e^{-S}$ and $X=(\tau,\bsl{x})$.
Here, $Z$ factors are chosen to only cancel the divergent part of the quantum corrections ($MS$ scheme\cite{Srednicki2007QFT}).
Since the divergence of the quantum corrections is given by $1/\epsilon^n$ with $n$ positive integer, the $Z$ factors must have the form
\begin{equation}
\label{eq:Z_series}
\ln(Z_{i})=\sum_{n=1}^{+\infty}\frac{A_i^{(n)}}{\epsilon^n}
\end{equation} with $i=\gamma,\phi,v_f,v_b,m,r,a,g,b,u$ and $``\ln "$ function formally defined as $\ln(1+x)=\sum_{n=1}^{+\infty}\frac{(-1)^{n-1}x^n}{n}$.
Now, we derive the expressions $\beta$ and $\Gamma$ functions, where we use $\Gamma$ instead of commonly used $\gamma$ to label the Gamma functions since the latter is reserved for the Majorana field.
Since $\widetilde{\mu}$ is not physical, the physical action as well as any physical observable should not depend on $\widetilde{\mu}$.
By defining
\begin{eqnarray}
\label{eq:S_Z_redef}
&&\phi^{(0)}=Z_{\phi}^{1/2} \phi,\ \gamma^{(0)}=Z_{\gamma}^{1/2} \gamma,\ g^{(0)}=\widetilde{\mu}^{\epsilon/2}Z_g g Z^{-1/2}_\phi Z^{-1}_\gamma,\nonumber\\
&&b^{(0)}=Z_b Z_\phi^{-3/2}b\widetilde{\mu}^{\epsilon/2},\ u^{(0)}=\widetilde{\mu}^{\epsilon}Z_u u Z^{-2}_\phi,\ v_f^{(0)}=Z_{v_f} v_f Z^{-1}_\gamma\nonumber\\
&& (v_b^{(0)})^2=Z_{v_b} v_b^2 Z^{-1}_\phi,\  m^{(0)}=Z_{m} m Z^{-1}_\gamma,\ r^{(0)}=Z_{r} r Z^{-1}_\phi,\nonumber\\
&&\text{and}\ a^{(0)}=Z_{\phi}^{-1/2} Z_a a \widetilde{\mu}^{-\epsilon/2},
\end{eqnarray}
the partition function becomes $Z=C_0 \int D\gamma^{(0)} D\phi^{(0)} e^{-S}$ with $S$ becomes
\begin{eqnarray}
\label{eq:S_phy}
&& S=\int d^d X \left\{\frac{1}{2}(\gamma^{(0)})^T[\partial_{\tau}+v_f^{(0)}(-i\bsl{\alpha}\cdot\bsl{\nabla})+m^{(0)}\sigma_y]\gamma^{(0)}\right.\nonumber\\
&& \left. +\frac{1}{2}g^{(0)}\phi^{(0)} (\gamma^{(0)})^T\sigma_y\gamma^{(0)}+\frac{1}{2}\phi^{(0)}(-\partial_\tau^2- (v_b^{(0)})^2\bsl{\nabla}^2)\phi^{(0)}\right.\nonumber\\
&& \left.+\frac{1}{2}r^{(0)}(\phi^{(0)})^2+a^{(0)}\phi^{(0)}+\frac{1}{3!}b^{(0)}(\phi^{(0)})^3+\frac{1}{4!}u^{(0)}(\phi^{(0)})^4\right\}\ .\nonumber\\
\end{eqnarray}
In this way, we obtain a partition function with an action that is independent of $\widetilde{\mu}$.
Since $\widetilde{\mu}$ is not physical, $S$ in the above equation should be the physical action that gives the physical observables.
It means the fields $\phi^{(0)}$ and $\gamma^{(0)}$ and the parameters $w^{(0)}$'s in the above equation are physical and independent of $\widetilde{\mu}$, where $w=v_f,v_b,r,m,a,g,b,u$.
As a result, we have
\begin{widetext}
\begin{eqnarray}
\label{eq:betaGamma}
&&\frac{d}{d\ln\widetilde{\mu}}\ln(\widetilde{\mu}^{\epsilon/2}Z_g g Z^{-1/2}_\phi Z^{-1}_\gamma)=0\Leftrightarrow \frac{d g}{d\ln\widetilde{\mu} }=-\frac{\epsilon}{2}g+\beta_g\ \text{with}\ \beta_g=g D(A^{(1)}_g-\frac{1}{2}A^{(1)}_\phi-A^{(1)}_\gamma)\nonumber\\
&&\frac{d}{d\ln\widetilde{\mu}}\ln(\widetilde{\mu}^{-\epsilon/2}Z_a a Z^{-1/2}_\phi)=0\Leftrightarrow \frac{d a}{d\ln\widetilde{\mu} }=\frac{\epsilon}{2}a+\beta_a\ \text{with}\ \beta_a=a D(A^{(1)}_a-\frac{1}{2}A^{(1)}_\phi)\nonumber\\
&&\frac{d}{d\ln\widetilde{\mu}}\ln(\widetilde{\mu}^{\epsilon/2}Z_b b Z^{-3/2}_\phi)=0\Leftrightarrow \frac{d b}{d\ln\widetilde{\mu} }=-\frac{\epsilon}{2}b+\beta_b\ \text{with}\ \beta_b=bD(A^{(1)}_b-\frac{3}{2}A^{(1)}_\phi)\nonumber\\
&&\frac{d}{d\ln\widetilde{\mu}}\ln(\widetilde{\mu}^{\epsilon}Z_u u Z^{-2}_\phi)=0\Leftrightarrow \frac{d u}{d\ln\widetilde{\mu} }=-\epsilon u +\beta_u\ \text{with}\ \beta_u=u D(A^{(1)}_u-2 A^{(1)}_\phi)\nonumber\\
&&\frac{d}{d\ln\widetilde{\mu}}\ln(Z_{v_f} v_f Z^{-1}_\gamma)=0\Leftrightarrow \frac{1}{v_f} \frac{d v_f}{d\ln\widetilde{\mu} }=\Gamma_{v_f}\ \text{with}\ \Gamma_{v_f}=D(A^{(1)}_{v_f}-A^{(1)}_\gamma)\nonumber\\
&&\frac{d}{d\ln\widetilde{\mu}}\ln(Z_{v_b} v_b^2 Z^{-1}_\phi)=0\Leftrightarrow \frac{1}{v_b} \frac{d v_b}{d\ln\widetilde{\mu} }=\Gamma_{v_b}\ \text{with}\ \Gamma_{v_b}=\frac{1}{2}D(A^{(1)}_{v_b}-A^{(1)}_\phi)\nonumber\\
&&\frac{d}{d\ln\widetilde{\mu}}\ln(Z_{m} m Z^{-1}_\gamma)=0\Leftrightarrow \frac{1}{m} \frac{d m}{d\ln\widetilde{\mu} }=\Gamma_{m}\ \text{with}\ \Gamma_{m}=D(A^{(1)}_{m}-A^{(1)}_\gamma)\nonumber\\
&&\frac{d}{d\ln\widetilde{\mu}}\ln(Z_{r} r Z^{-1}_\phi)=0\Leftrightarrow \frac{1}{r} \frac{d r}{d\ln\widetilde{\mu} }=\Gamma_{r}\ \text{with}\ \Gamma_{r}=D(A^{(1)}_{r}-A^{(1)}_\phi)\nonumber\\
&&\Gamma_\phi=\frac{1}{2}\frac{d}{d\ln\widetilde{\mu}}\ln(Z_{\phi})=-\frac{1}{2}D(A^{(1)}_\phi)\nonumber\\
&&\Gamma_\gamma=\frac{1}{2}\frac{d}{d\ln\widetilde{\mu}}\ln(Z_{\gamma})=-\frac{1}{2}D(A^{(1)}_\gamma)\ ,
\end{eqnarray}
\end{widetext}
where $D=\frac{g}{2}\partial_g+\frac{b}{2}\partial_b+u\partial_u-\frac{a}{2} \partial_a$ and the condition that $\beta$ and $\Gamma$ functions are finite at $\epsilon\rightarrow 0$ is used.
Now we derive the Callan-Symanzik equation.
The physical $N$-point function is defined as
\begin{equation}
F^{(0)}_{N_\phi,N_\gamma}(X_n,w^{(0)})=\left\langle \left(\prod_{n_1 = 1}^{N_\phi} \phi^{(0)}_{X_{n_1}}\right)\left(\prod_{n_2 = 1}^{N_\gamma} \gamma^{(0)}_{X_{n_2},i_{n_2}}\right)\right\rangle
\end{equation}
with $X_n$ indicating all its coordinate dependence.
The above equation is related with the $N$-point function of $\gamma$ and $\phi$ by \eqnref{eq:S_Z_redef}:
\eqn{
&& F_{N_\phi,N_\gamma}(X_n,w,\widetilde{\mu})=\left\langle \left(\prod_{n_1 = 1}^{N_\phi} \phi_{X_{n_1}}\right)\left(\prod_{n_2 = 1}^{N_\gamma} \gamma_{X_{n_2},i_{n_2}}\right)\right\rangle\nonumber\\
&&=Z_{\phi}^{-N_{\phi}/2}Z_{\gamma}^{-N_{\gamma}/2}F^{(0)}_{N_\phi,N_\gamma}\ .
}
Combining the fact that $F^{(0)}_{N_\phi,N_\gamma}$ is independent of $\widetilde{\mu}$ and \eqnref{eq:betaGamma}, we arrive at the Callan-Symanzik equation with respect to $\widetilde{\mu}$:
\eqn{
\label{eq:CS_mu}
&& (\frac{\partial}{\partial \ln\widetilde{\mu}}+N_{\phi}\Gamma_{\phi}+N_\gamma \Gamma_\gamma+ \beta_a \partial_a + \beta_g \partial _g+\beta_u \partial_u+\beta_b\partial_b\nonumber\\
&&+v_f\Gamma_{v_f}\partial_{v_f}+v_b\Gamma_{v_b}\partial_{v_b}+m\Gamma_m\partial_m+r\Gamma_r\partial_r-\epsilon D)\nonumber\\
&&F_{N_\phi,N_\gamma}(X_n,w,\widetilde{\mu})=0\ .
}
But we want the Callan-Symanzik equation with respect to the physical scale instead of the non-physical $\widetilde{\mu}$.
To do so, consider the $N$-point function with scaled coordinates $F_{N_\phi,N_\gamma}(t X_n,w,\widetilde{\mu})$.
By defining $\gamma_{X}' =\gamma_{tX}t^{[\gamma]}$, $\phi_{X}'=\phi_{tX} t^{[\phi]}$, $w'=w t^{[w]}$ and $\widetilde{\mu}'=\widetilde{\mu} t$ , we have
\begin{equation}
F_{N_\phi,N_\gamma}(t X_n,w,\widetilde{\mu})=t^{-N_\phi [\phi]-N_\gamma [\gamma]}F_{N_\phi,N_\gamma}(X_n,w',\widetilde{\mu}')\ .
\end{equation}
Differentiating the above equation by $t$ at $t\rightarrow 1$ and using \eqnref{eq:CS_mu}, we have
\begin{eqnarray}
&& X_n\partial_{X_n}F_{N_{\phi}, N_{\gamma}}(X_n,w,\widetilde{\mu})
=
\left[
-N_\phi (\frac{d-2}{2}+\Gamma_\phi)\right.\nonumber\\
&&
\left.-N_\gamma(\frac{d-1}{2}+\Gamma_\gamma)+((3-\frac{\epsilon}{2})a-\beta_a)+(\frac{\epsilon}{2}g-\beta_g)\partial_g\right.\nonumber\\
&&
\left.+(\frac{\epsilon}{2}b+b-\beta_b)\partial_b+(\epsilon u -\beta_u)\partial_u+(-v_f \Gamma_{v_f})\partial_{v_f}\right.\nonumber\\
&&
\left.+(-v_b\Gamma_{v_b})\partial_{v_b}+(1-\Gamma_m)m\partial_m+(2-\Gamma_r)r\partial_r
\right]\nonumber\\
&&
F_{N_{\phi}, N_{\gamma}}(X_n,w,\widetilde{\mu})\ .
\end{eqnarray}
The meaning of the above equation can be better illutrated in the integrated form:
\eqn{
\label{eq:F_l}
&& F_{N_{\phi}, N_{\gamma}}(e^l X_n,w(0),\widetilde{\mu})
=
[\zeta_{\phi}(l)]^{N_\phi}[\zeta_{\gamma}(l)]^{N_\gamma}\nonumber\\
&&F_{N_{\phi}, N_{\gamma}}( X_n,w(l),\widetilde{\mu})\ ,
}
where
\begin{eqnarray}
\label{eq:RG_gen}
&&\frac{dg}{dl}=\frac{\epsilon}{2}g-\beta_g,\ \frac{db}{dl}=(\frac{\epsilon}{2}+1)b-\beta_b,\ \frac{du}{dl}=\epsilon u -\beta_u\nonumber\\
&& \frac{dv_f}{dl}=-v_f \Gamma_{v_f},\ \frac{dv_b}{dl}=-v_b\Gamma_{v_b},\
\frac{dm}{dl}=(1-\Gamma_m)m\nonumber\\
&&\frac{dr}{dl}=(2-\Gamma_r)r,\ \frac{d a}{d l}=(3-\frac{\epsilon}{2})a-\beta_a\\
&& \frac{d\zeta_\phi}{dl}=-(\frac{d-2}{2}+\Gamma_\phi)\zeta_\phi,\ \frac{d\zeta_\gamma}{dl}=-(\frac{d-1}{2}+\Gamma_\gamma)\zeta_\gamma\ ,\nonumber
\end{eqnarray}
and $\zeta_\phi(0)=\zeta_\gamma(0)=1$.
\eqnref{eq:F_l} indicates that the $N$-point function at a larger scale $e^l X_n$ is the same as the $N$-point function at $X_n$ with scaled fields $\phi,\gamma$ and parameters $w$ according to \eqnref{eq:RG_gen}.
It means, we can define an action $S^l$ that is the same as $S$ except that the parameters $w$ in $S^l$ can scale with $l$ according to \eqnref{eq:RG_gen}.
In this case, the N-point function generated by $S^l$ at $X_n$ and the N-point function given by $S$ at $e^l X_n$ just deviate from each other by a factor $[\zeta_{\phi}(l)]^{N_\phi}[\zeta_{\gamma}(l)]^{N_\gamma}$.
Then, $S_l$ can be viewed as an effective action of $S$ at a larger scale $e^l$.
And \eqnref{eq:RG_gen} is called the RG equations.
In the following, we will derive the RG equations to the one-loop order.

\subsection{One-Loop RG Equations}

\begin{figure}[t]
\includegraphics[width=\columnwidth]{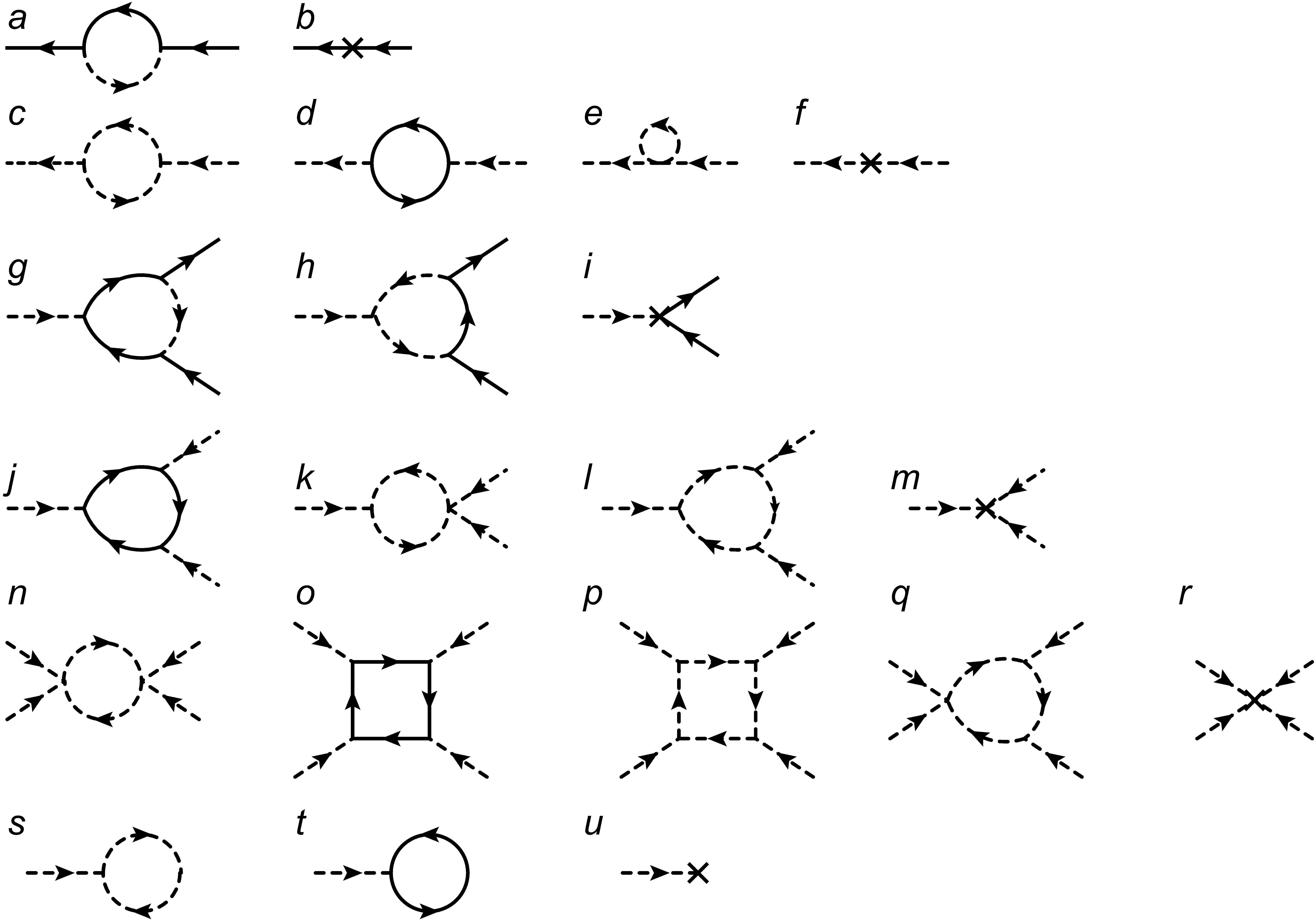}
\caption{\label{fig:one-loop-dia}
One-loop Feynman diagrams and the corresponding counter-terms for the $Z$ factors in \eqnref{eq:S_Z}.
The solid(dashed) line is the fermion(boson) propagator in \eqnref{eq:G_gamma} (\eqnref{eq:G_phi}).
The cross maker stands for the counter-term in \eqnref{eq:S_ct}.
}
\end{figure}

In order to obtain  \eqnref{eq:RG_v},\eqref{eq:RG_gu}, \eqref{eq:RG_mb} and \eqref{eq:RG_ra}, we need to find the corresponding $\beta$ and $\Gamma$ functions in \eqnref{eq:RG_gen}.
According to \eqnref{eq:betaGamma} and \eqnref{eq:Z_series}, it is equivalent to deriving the expression of the $Z$ factors in \eqnref{eq:S_Z}.
Since the $Z$ factors are given by the quantum corrections, we need to evaluate the loop diagrams.
For convenience, we seperate \eqnref{eq:S_Z} into two parts
$S=S_c+S_{ct}$ with
\begin{eqnarray}
\label{eq:S_c}
&&S_c=\int d^d X \left\{
\frac{1}{2}\gamma^T[ \partial_{\tau}+ v_f(-i\bsl{\alpha}\cdot\bsl{\nabla})+ m\sigma_y]\gamma\right.\nonumber\\
&&\left.+\frac{1}{2} g\widetilde{\mu}^{\epsilon/2}\phi \gamma^T\sigma_y\gamma+\frac{1}{2}\phi(- \partial_\tau^2-v_b^2\bsl{\nabla}^2+r)\phi\right.\nonumber\\
&&\left. +a\widetilde{\mu}^{-\epsilon/2}\phi+\frac{1}{3!} b\widetilde{\mu}^{\epsilon/2}\phi^3+\frac{1}{4!}u\widetilde{\mu}^{\epsilon}\phi^4\right\}
\end{eqnarray}
and
\begin{eqnarray}
\label{eq:S_ct}
&&S_{ct}=\int d^d X \left\{
\frac{1}{2}\gamma^T[(Z_\gamma-1) \partial_{\tau}+(Z_{v_f}-1) v_f(-i\bsl{\alpha}\cdot\bsl{\nabla})\right.\nonumber\\
&&\left.+(Z_m-1) m\sigma_y]\gamma+\frac{1}{2}(Z_g-1) g\widetilde{\mu}^{\epsilon/2}\phi \gamma^T\sigma_y\gamma\right.\\
&&\left.+\frac{1}{2}\phi[-(Z_\phi-1) \partial_\tau^2-(Z_{v_b}-1) v_b^2\bsl{\nabla}^2+(Z_r-1) r]\phi\right.\nonumber\\
&&\left.
+(Z_a-1)a\widetilde{\mu}^{\epsilon/2} \phi+\frac{1}{3!}(Z_b-1) b\widetilde{\mu}^{\epsilon/2}\phi^3+\frac{1}{4!}(Z_u-1) u\widetilde{\mu}^{\epsilon}\phi^4\right\}
\ .\nonumber
\end{eqnarray}
Here $S_c$ is \eqnref{eq:S_Z} without quantum corrections and $S_{ct}$ is called the counter-terms.
According to \eqnref{eq:S_c}, the fermion propagator is
\begin{equation}
\label{eq:G_gamma}
G_\gamma(k)=(i\omega-m\sigma_y-v_f \bsl{k}\cdot\bsl{\alpha})^{-1}
\end{equation}
and the boson propagator reads
\begin{equation}
\label{eq:G_phi}
G_\phi(q)=(\nu^2+v_b^2 \bsl{q}^2 + r)^{-1}\ .
\end{equation}
Here $k=(\omega,\bsl{k})$ and $q=(\nu,\bsl{q})$.
Equiped with \eqnref{eq:S_c}-\eqref{eq:G_phi}, we can evaluate the loop diagrams.
For simplicity, we only consider the one-loop diagrams, which together with the corresponding counter-terms are shown in \figref{fig:one-loop-dia}.

The one-loop contribution to the fermion self-energy $\Sigma(k)$ is given by \figref{fig:one-loop-dia} a and b, which reads
\begin{eqnarray}
\label{eq:f_pro}
&&\Sigma(k)=g^2\widetilde{\mu}^{\epsilon}\int \frac{d^d q }{(2\pi)^d}\sigma_y G_\gamma(k-q)\sigma_y G_\phi(q)\nonumber\\
&&+\left[-(Z_\gamma-1)i\omega+(Z_m-1)m\sigma_y+(Z_{v_f}-1)v_f \bsl{\alpha}\cdot\bsl{k}\right]\\
&&=-\frac{g^2}{4 \pi ^2 v_{b} (v_{b}+v_{f})^2}\frac{1}{\epsilon}\left[i\omega+\frac{v_{b}+v_{f}}{v_{f}}m\sigma_y-\frac{2 v_{b}+v_{f}}{3 v_{f}}v_f \bsl{\alpha}\cdot\bsl{k}\right]\nonumber\\
&&+\left[-(Z_\gamma-1)i\omega+(Z_m-1)m\sigma_y+(Z_{v_f}-1)v_f \bsl{\alpha}\cdot\bsl{k}\right]+O(\epsilon^0)\nonumber
\ ,
\end{eqnarray}
where we only keep the divergent terms since $Z$ factors are chosen to only cancel the divergent part.

The one-loop contribution to the boson self-energy $\Pi(q)$ is given by \figref{fig:one-loop-dia} c, d, e and f, which reads
\begin{eqnarray}
\label{eq:b_pro}
&&\Pi(q)=\frac{b^2\widetilde{\mu}^{\epsilon}}{2}\int \frac{d^d q_1}{(2\pi)^d}G_\phi(q_1-q)G_\phi(q_1)\nonumber\\
&&-\frac{g^2\widetilde{\mu}^{\epsilon}}{2}\int \frac{d^d k}{(2\pi)^2}\text{Tr}\left[G_\gamma(k)\sigma_y G_\gamma(k+q)\sigma_y\right]\nonumber\\
&&-\frac{u \widetilde{\mu}^{\epsilon}}{2}\int \frac{d^d q_1}{(2\pi)^d}G_\phi(q_1)\nonumber\\
&&-\left[(Z_\phi-1)\nu^2+(Z_{v_b}-1)v_b^2 \bsl{q}^2+(Z_r-1)r \right]\\
&&=\frac{b^2}{\left(16 \pi ^2 v_{b}^3\right) \epsilon }-\frac{g^2 \left(6 m^2+\nu ^2+v_{f}^2 \bsl{q}^2\right)}{16 \pi ^2 v_{f}^3 \epsilon }+\frac{r u}{\left(16 \pi ^2 v_{b}^3\right) \epsilon }\nonumber\\
&&-\left[(Z_\phi-1)\nu^2+(Z_{v_b}-1)v_b^2 \bsl{q}^2+(Z_r-1)r \right]+O(\epsilon^0)\ .\nonumber
\end{eqnarray}

The fermion-boson coupling term $\Gamma^{\phi\gamma\gamma}$ has the one-loop correction given by \figref{fig:one-loop-dia} g,h and i, which reads
\begin{eqnarray}
\label{eq:g}
&&F^{\phi\gamma\gamma}=-g^3 \widetilde{\mu}^{3\epsilon/2}\int \frac{d^d q'}{(2\pi)^d}[\sigma_y G_\gamma(q')\sigma_y G_\gamma(q')\sigma_y ]G_\phi(q')\nonumber\\
&&-g \widetilde{\mu}^{\epsilon/2} (Z_g-1)\sigma_y+O(\epsilon^0)\nonumber\\
&&=
\frac{g^3}{4 \pi ^2 v_{b} v_{f} \epsilon  (v_{b}+v_{f})}
\sigma_y -g (Z_g-1)\sigma_y+O(\epsilon^0)\ ,
\end{eqnarray}
where the contribution of \figref{fig:one-loop-dia}h is not explicitly included since it is not divergent for $d=4-\epsilon$.

For the $\phi^3$ term, the one-loop contribution is given by \figref{fig:one-loop-dia} j, k, l and m, and reads
\begin{eqnarray}
\label{eq:b}
&&F^{\phi^3}=-g^3 \widetilde{\mu}^{3\epsilon/2} \int \frac{d ^d k}{(2\pi)^d}\text{Tr}[\sigma_y G_\gamma(k)\sigma_y G_\gamma(k)\sigma_y G_\gamma(k)]\nonumber\\
&&+\frac{3}{2} u b \widetilde{\mu}^{3\epsilon/2} \int \frac{d ^d q}{(2\pi)^d} G_\phi(q)^2 -(Z_b - 1) b \widetilde{\mu}^{\epsilon/2}+O(\epsilon^0)\nonumber\\
&&
=-\frac{3 g^3 m}{4 \pi ^2 v_{f}^3 \epsilon }+\frac{3 b u}{\left(16 \pi ^2 v_{b}^3\right) \epsilon }-(Z_b - 1) b +O(\epsilon^0)\ ,
\end{eqnarray}
where \figref{fig:one-loop-dia}l is not divergent and included in $O(\epsilon^0)$.

\figref{fig:one-loop-dia} n, o, p, q and r give the one-loop contribution of the $\phi^4$ term.
However, only \figref{fig:one-loop-dia} n, o and r are divergent and contribute to the RG equations, which gives
\begin{eqnarray}
\label{eq:u}
F^{\phi^4}&&= \frac{3 u^2}{2}\widetilde{\mu}^{2\epsilon}\int \frac{d^d q}{(2\pi)^d}G_\phi(q)^2\nonumber\\
&&-3g^4 \widetilde{\mu}^{2\epsilon}\int \frac{d^d k}{(2\pi)^d}\text{Tr}[G_\gamma(k)\sigma_yG_\gamma(k)\sigma_yG_\gamma(k)\sigma_yG_\gamma(k)\sigma_y]\nonumber\\
&&-(Z_u-1)u\widetilde{\mu}^{\epsilon}+O(\epsilon^0)\nonumber\\
&&=\frac{3 u^2}{16 \pi^2 v_b^3}\frac{1}{\epsilon}-\frac{3 g^4}{4 \pi^2 v_f^3}\frac{1}{\epsilon}-(Z_u-1)u+O(\epsilon^0)\ .
\end{eqnarray}

\figref{fig:one-loop-dia} s, t and u give the one-loop contribution of the $\phi$ term, which gives
\begin{eqnarray}
\label{eq:a}
&&F^{\phi}= -\frac{g\widetilde{\mu}^{\epsilon/2}}{2}\int \frac{d^d k}{(2\pi)^d}\Tr[\sigma_y G_\gamma(k)]- \frac{b\widetilde{\mu}^{\epsilon/2}}{2}\int \frac{d^d q}{(2\pi)^d} G_\phi(q)\nonumber\\
&&-(Z_a-1)a\widetilde{\mu}^{-\epsilon/2}=\frac{-g m^3}{8\pi^2 v_f^3 \epsilon}+\frac{r b}{16 \pi^2 v_b^3 \epsilon}-(Z_a-1)a+O(\epsilon^0)\ .\nonumber\\
\end{eqnarray}

According to \eqnref{eq:f_pro}-\eqref{eq:a} and \eqnref{eq:Z_series}, we have the expressions of $A^{(1)}_{a}$'s as
\begin{eqnarray}
\label{eq:A_exp}
&& A_\gamma^{(1)}=-\frac{g^2}{4 \pi ^2 v_{b} (v_{b}+v_{f})^2}\nonumber\\
&& A_{v_f}^{(1)}=-\frac{g^2 (2 v_{b}+v_{f})}{12 \pi ^2 v_{b} v_{f} (v_{b}+v_{f})^2}\nonumber\\
&& A_{m}^{(1)}=\frac{g^2}{4 \pi ^2 v_{b} v_{f} (v_{b}+v_{f})}\nonumber\\
&& A_{\phi}^{(1)}=-\frac{g^2}{16 \pi ^2 v_{f}^3}\nonumber\\
&& A_{v_b}^{(1)}=-\frac{g^2}{16 \pi ^2 v_{b}^2 v_{f}}\nonumber\\
&& A_{g}^{(1)}=
\frac{g^2}{4 \pi ^2  v_{b} v_{f}   (v_{b}+v_{f})}
\nonumber\\
&& A_{b}^{(1)}=\frac{3 \left(\frac{u}{v_{b}^3}-\frac{4 g^3 m}{b v_{f}^3}\right)}{16 \pi ^2}\nonumber\\
&& A_{u}^{(1)}=\frac{3 \left(u^2 v_{f}^3-4 g^4 v_{b}^3\right)}{16 \pi ^2 u v_{b}^3 v_{f}^3}\nonumber\\
&& A_{r}^{(1)}=\frac{-6 m^2 g^2 v_b^3+(u r+b^2) v_f^3}{16 \pi ^2 v_b^3 v_f^3 r}\nonumber\\
&& A_{a}^{(1)}=\frac{v_f^3 b r -2 g m^3 v_b^3}{16 \pi ^2 v_{b}^3 v_{f}^3 a} .
\end{eqnarray}
Plug the above equation into \eqnref{eq:betaGamma}, we can get the $\beta$ and $\Gamma$ functions and then we can derive the RG equations according to \eqnref{eq:RG_gen}.
Due to the transformation \eqnref{eq:gub2mu}, $a$, $g$, $u$ and $b$ in the obtained RG equations should be replaced by $\widetilde{a}$, $\widetilde{g}$, $\widetilde{u}$ and $\widetilde{b}$ according to the convention defined in the main text.
As a result, we can get  \eqnref{eq:RG_v},\eqref{eq:RG_gu}, \eqref{eq:RG_mb} and \eqref{eq:RG_ra} in the main text.

From  \eqnref{eq:RG_v},\eqref{eq:RG_gu}, \eqref{eq:RG_mb} and \eqref{eq:RG_ra} in the main text, we discuss the RG equations of $w=v_f/v_b$ and the RG equation of $u/g^2$.
The RG equation of $w$ reads
\eq{
\frac{d w}{d l}=-\frac{g^2 (w-1) (w (w (3 w+25)+9)+3)}{96 \pi ^2 v_b^3 w^2 (w+1)^2}\ ,
}
which can be re-written as
\eq{
\frac{d \widetilde{w}}{d l}=-\frac{g^2 \widetilde{w} (\widetilde{w} (\widetilde{w} (3 \widetilde{w}+34)+68)+40)}{96 \pi ^2 v_{b}^3 (\widetilde{w}+1)^2 (\widetilde{w}+2)^2}=-\frac{5 g^2 \widetilde{w}}{48 \pi ^2 v_{b}^3}+O\left(\widetilde{w}^2\right)
}
with $\widetilde{w}=w-1$.
The RG equation of $\widetilde{w}$ shows that the velocity ratio $v_f/v_b$ flows exponentially to 1 when it is close to 1.
On the other hand, the RG equation of $u/g^2$ for $v_f=v_b=1$ reads
\eq{
\frac{d (u/g^2)}{d l}=-\frac{g^2}{16\pi^2}(\frac{u}{g^2}-3)(3 \frac{u}{g^2}+4)\ .
}
Define $y=u/g^2-3$, then the above equation can be re-written as
\eq{
\frac{d y}{d l}=-\frac{g^2}{16\pi^2}y(3y+13)\ ,
}
where the existence of y-linear term indicates the exponential flow of $u/g^2$ to $3$.
Moreover, according to Eq.\,(5) of the main text, the RG equation of $z=b/(m g)-3$ for $v_f=v_b=1$ and $u/g^2=3$ is linear in $z$:
\eq{
\frac{d z}{d l}=-\frac{g^2}{4 \pi^2}z\ ,
}
indicating the exponential flow of $b/(m g)$ to $3$.
Therefore, although the classical dimensions of $v_f/v_b$, $u/g^2$ and $b/(m g)$ are zero, their stable flows to the SUSY hypersurface are exponential instead of logarithmic after including the quantum correction.

\subsection{Higher-loop Contribution}
Unlike the continuous phase transition, the validity of perturbation theory in series of loops is not obvious in our case due to the existence of relevant $m$, $r$ and $b$, which might make the higher-loop terms not small.
In this part, we discuss the validity of neglecting the higher-loop terms.
Here we still replace the $\tilde{a}, \tilde{g},\tilde{u},\tilde{b}$ in the main text by $a,g,u,b$ as above, and we only consider the 1PI connected graphs with loops, which do not contain $a$.

As shown above, the RG equations are obtained from the counter-terms.
The counter-terms are determined by the graphs with non-negative superficial degree of divergence $D=4 L-2 I_\phi-I_\gamma$, where $L$ is the number of loops in the graph, and $I_\phi$ and $I_\gamma$ are the numbers of internal bosonic and fermionic propagators, respectively.\cite{Srednicki2007QFT}
To estimate the order of higher-loop terms, we need to know the structure of a generic connected graph.
As one end of the external propagator and the two ends of the internal propagator are connected to vertexes, we have $E_\phi+2I_\phi=V_g+3 V_b+4 V_u$ and $E_\gamma+2I_\gamma=2V_g$, where $E_\phi$ and $E_\gamma$ are the numbers of external bosonic and fermionic propagators, respectively, and $V_g$, $V_b$ and $V_u$ are number of $g$, $b$ and $u$ vertexes, respectively.
Furthermore, the momentum conservation gives $L=I_\phi+I_\gamma-(V_g+V_b+V_u)+1$.
From the above relations, we have
\eqn{
\label{eq:LD}
&&L=\frac{-E_\phi-E_\gamma+V_g+V_b+2 V_u+2}{2}\nonumber\\
&&D=4-E_\phi-\frac{3 E_\gamma}{2} - V_b\ .
}
Equipped with those relations, we next address the possible issues brought by the three relevant parameters by estimating the order of the $L$-loop contribution when $a\sim m^3/g$, $u\sim g^2$, $b\sim g m$, $r\sim m^2$ and $v_f\sim v_b\sim 1$, which is near the SUSY hypersurface.
With the assumption, we also have $m'\sim m$.

Since only $D\geq 0$ graphs contribute to counter-terms, high powers of $b$ do not exist in the RG equations as they can make $D$ negative according to \eqnref{eq:LD}.
Therefore, the relevant $b$ cannot cause any divergence in summing the series.
On the other hand, $D\geq 0$ requires $2 E_\phi+3 E_\gamma\leq 8$, leading to the following six combinations: $(E_\phi,E_\gamma)=(1,0)$, $(2,0)$, $(3,0)$, $(4,0)$, $(0,2)$ or $(1,2)$.
Here we use the fact that $E_\gamma$ can only be even.
Clearly, those six combinations correspond exactly to those counter-terms that we include in \eqnref{eq:S_Z}, verifying the renormalizability of our theory.
In the following, we estimate the order of the counter-terms for the six combinations, namely the $Z$ factors.
A generic connected graph with loop has the form $g^{V_g} b^{V_b} u^{V_u} F(k)$ with the loop-integral part $F(k)$ having dimension $[F(k)]=D$, where $\widetilde{\mu}$ is omitted since $\epsilon\rightarrow 0$ and will be taken into account in the $1/\epsilon$ expansion in the following.
For the condition that we choose, only $m$ and $k$ can carry dimensions, and thus the graph is approximately $g^{2 L-2 + E_\phi+E_\gamma} \sum_{n} C_n m^{D+V_b-n} k^n$, where $C_n$ is dimensionless and given by the loop integral.
Based on this estimation, all the $Z$ factors approximately have the form $1+\sum_{L=1}^\infty C_L (g^2)^L$, where the dimensionless $C_L$ has the form $\sum_{i=1}^L C^{(i)}_L/\epsilon^i$ and only depends on $\ln(|m|/\tilde{\mu})$.
Since only $C^{(1)}_L$ contributes to the RG equations, we only need to care about whether $\sum_{L=1}^\infty C_L ^{(1)}(g^2)^L$ is well defined.
As $m$ is relevant, $|m|\gg \tilde{\mu}$ is commonly true at a relatively large scale, and thus the $C^{(1)}_L\sim [\ln(|m|/\tilde{\mu})]^{L-1}$.
Then, the $L$-loop term in the series of interest is of the order $\sim [\ln(|m|/\tilde{\mu})]^{L-1}(g^2)^L$.
Therefore, the higher-loop contributions to the RG equations are neglectable if $\ln(|m|/\tilde{\mu})< 1/g^2\sim 1/\epsilon\sim \ln(\Lambda/\tilde{\mu})$, where $\Lambda$ is the ultraviolet energy cut-off of the model.
Furthermore, it means $\ln(|m'|/\tilde{\mu})< \ln(\Lambda/\tilde{\mu})$.
Clearly, in the assumption that the ultraviolet energy cut-off is much larger than any dimensionful parameter of the model (or similarly $l<l_c$), the higher-loop terms can be neglected and the one-loop result is trustworthy.

\subsection{Emergent-SUSY Region}

\begin{figure}[t]
\includegraphics[width=\columnwidth]{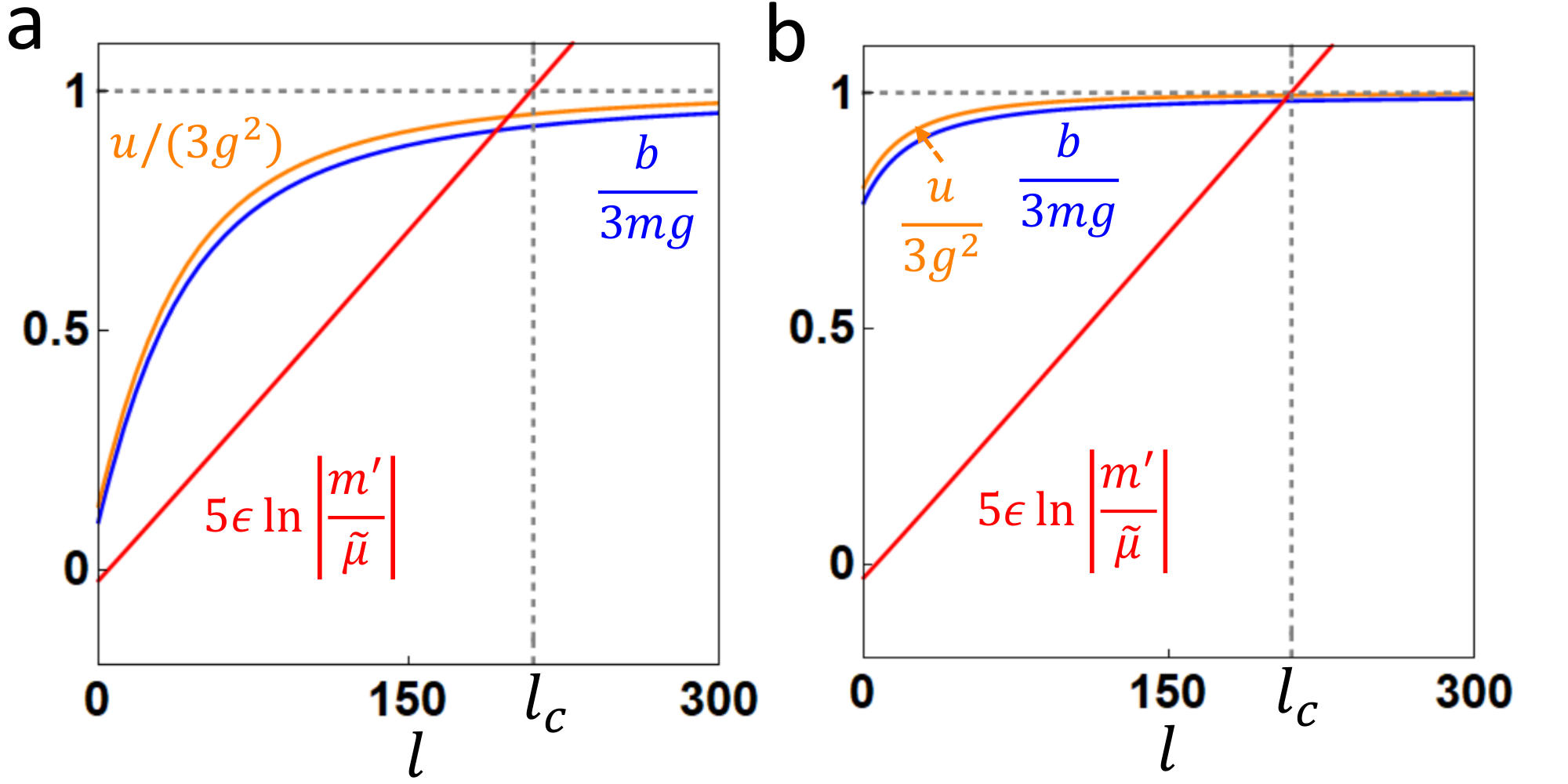}
\caption{\label{fig:SUSY_reg}
The graph shows how $u/(3 g^2)$ (orange), $b/(3 m g)$ (blue) and $5 \epsilon \ln(|m'/\widetilde{\mu}|)$ (red) change with the scale $l$.
(a) and (b) are examples of paths 1 and 2 in \figref{fig:RG_scheme}d, respectively. The horizontal gray dashed line is at $1$, and the vertical gray dashed line is at $l=l_c$.
}
\end{figure}

In this part, we discuss the example of paths 1 and 2 of \figref{fig:RG_scheme}d, as shown in \figref{fig:SUSY_reg}(a) and (b), respectively.
\figref{fig:SUSY_reg} is ploted according to \eqnref{eq:RG_gu}, \eqnref{eq:RG_mb} and \eqnref{eq:RG_ra} with $a= \frac{m}{g}(r-m^2)$ fixed.
$m'$ is determined by the relation for \eqnref{eq:S_vshift}, and the critical scale is defined as $\ln(|m'/\widetilde{\mu}|)=\frac{1}{5}\frac{1}{\epsilon}$ since $1/\epsilon\sim \log(\Lambda/\widetilde{\mu})$.
The values of parameters are $\widetilde{g}_i=0.8, \epsilon=10^{-3}, m_i/\widetilde{\mu}=0.1, r_i/\widetilde{\mu}^2=0.1$ for both subgraphs of \figref{fig:SUSY_reg}, while  $(u_i/g_i^2,b_i/(m_i g_i))=(0.4,0.3)$ for (a) and $(u_i/g_i^2,b_i/(m_i g_i))=(2.4,2.3)$ for (b).
Here the subscript ``$i$" means they are initial (at $l=0$) values.
According to \figref{fig:SUSY_reg}, path 1 cannot reach the SUSY point before $l$ reaches $l_c$, while path 2 can be very close to the SUSY point.

\section{First-Order Phase Transitions}
\label{app:FOQPT}
In this section, we show details on determining the phases of \eqnref{eq:S*} with $a^*=0$.
For convenience, we drop the ``$*$'' of all parameters in \eqnref{eq:S*}.

$\langle \phi \rangle$ is given by the global minimum of bosonic part of \eqnref{eq:S*} with $a^*=0$, which reads
\begin{equation}
S_b^*=\int d\tau d^{d-1}x\left[\frac{1}{2}\phi(-\partial^2+r)\phi+\frac{1}{2}g m\phi^3+\frac{1}{8}g^2\phi^4\right]\ .
\end{equation}
Since $\langle \phi \rangle$ indicates the macroscopic magnetic ordering and must be real, we should impose $\phi^\dagger_{\tau,\bsl{x}}=\phi_{\tau,\bsl{x}}$ when solving for the global minimum of $S^*_b$.
$S_b^*$ can be splitted into two parts $S_b^*=S_T+S_V$ with
\begin{equation}
S_T=\int d\tau d^{d-1}x \frac{1}{2}\phi(-\partial^2)\phi\ \text{and}\ S_V=\int d\tau d^{d-1}x\ V(\phi)\ .
\end{equation}
Here $V(\phi)=\frac{1}{2}r\phi^2+\frac{1}{2}g m\phi^3+\frac{1}{8}g^2\phi^4$.
By defining $\phi_{\tau,\bsl{x}}=\frac{1}{\beta\mathcal{V}}\sum_{\omega,\bsl{k}}e^{-i\omega \tau+i\bsl{k}\cdot\bsl{x}}\phi_{\omega,\bsl{k}}$ and $\phi_{\omega,\bsl{k}}^*=\phi_{-\omega,-\bsl{k}}$, $S_T$ can be re-written as
\begin{equation}
S_T=\frac{1}{\beta\mathcal{V}}\sum_{\omega,\bsl{k}}(\omega^2+\bsl{k}^2)|\phi_{\omega,\bsl{k}}|^2\ .
\end{equation}
It means $S_T\geq 0$ and $S_T=0$ holds only if $\phi$ is uniform in $(\tau,\bsl{x})$.
On the other hand, for any non-uniform field $\phi_{\tau,\bsl{x}}$, we can always find a uniform field $\phi^{(0)}$ such that $S_{V}[\phi^{(0)}]\leq S_{V}[\phi]$.
(Simply, one can pick one position $(\tau_0,\bsl{x}_0)$ such that $V(\phi_{\tau,\bsl{x}})\geq V(\phi_{\tau_0,\bsl{x}_0})$ holds for any $(\tau,\bsl{x})$ and define the uniform field as $\phi^{(0)}=\phi_{\tau_0,\bsl{x}_0}$.)
Therefore, the global minimum of $S^*_b$ must be uniform given $\phi^\dagger_{\tau,\bsl{x}}=\phi_{\tau,\bsl{x}}$.
Therefore, we only need to minimize the bosonic potential $V(\phi)$ to obtain the minimum field.
The extrema of $V(\phi)$ are given by
\begin{equation}
\frac{dV(\phi)}{d\phi}=\frac{1}{2} \phi  \left(g^2 \phi ^2+3 g m \phi +2 r\right)=0\ .
\end{equation}
In this case, if $9 m^2 < 8 r$, there is only one extremum that is $\phi_0=0$, and if $9 m^2 \geq 8 r$, the extrema are
\begin{equation}
\phi_0=0\ \text{and}\ \phi_\pm=\frac{-3 m \pm\sqrt{9 m^2-8 r}}{2 g}\ .
\end{equation}
Clearly, $\phi_\pm\neq 0$ as long as $r\neq 0$ and they exist.
The values of $V(\phi)$ at the three extrema are
\begin{eqnarray}
V(\phi_0)=0\ \text{and}\ V(\phi_\pm)=\frac{1}{16}\phi^2_\pm(2 g m \phi_\pm+4r)\ .
\end{eqnarray}
Now we discuss the global minimum of $V(\phi)$.
As mentioned before, if $9 m^2 < 8 r$, $\phi_0$ is only one extremum and thereby is the global minimum.
If $9 m^2/8 \geq r >m^2$, we have $4r-3m^2>m^2>|m\sqrt{9m^2-8r}|$ , which gives $2 g m \phi_\pm +4 r=4 r-3m^2\pm m\sqrt{9m^2-8r}>0$ and thus $V(\phi_\pm)>0$.
Therefore, $\phi_0=0$ is the global minimum if $r>m^2$.
For $r<m^2$ and $m>0$, we have
\begin{equation}
2 g m \phi_-+4r< m^2-m\sqrt{9m^2-8r}<0\ ,
\end{equation}
which gives $V(\phi_-)<0$.
$m>0$ results in $2 g m \phi_-+4 r<2 g m \phi_++4 r$ and $\phi_-^2>\phi_+^2$.
In this case, if $2 g m \phi_+ + 4 r\geq 0$, $V(\phi_+)\geq 0> V(\phi_-)$.
And if $2 g m \phi_+ - 4 r< 0$, $\phi_-^2>\phi_+^2$ gives $V(\phi_+)> V(\phi_-)$.
Then, we have $V(\phi_-)<V(\phi_+)$.
Therefore, $\phi_-$ is the global minimum if $r<m^2$ and $m<0$.
Since $V(\phi)$ is invariant under $m\rightarrow -m$ and $\phi\rightarrow -\phi$, the global minimum is at $\phi_+$ if $r<m^2$ and $m>0$.
Therefore, $r=m^2, m\neq 0$ and $r<0,m=0$ are where the first order phase transition happens.
In the following, we show the form of the action at the first order phase transition.

At $r=m^2$ with $m\neq 0$, there are two degenerate vacua: (i) $\langle \phi \rangle=0$, and (ii) $\langle \phi \rangle=-2m/g$.
Around the first vacuum, the form of the action is just $S^*$ with $r=m^2$ and $a=0$.
Around the second vacuum, we need to use the fluctuation $\delta \phi= \phi-(-2m/g)$ in $S^*$ with $a=0$, and the resulted action for $\gamma$ and $\delta\phi$ reads
\eqn{
\label{eq:S_msq_v2}
&& S^*=\int d\tau d^{d-1}x \left\{
\frac{1}{2}\gamma^T(i\partial_{\mu}\alpha^{\mu}-m\sigma_y)\gamma+\frac{1}{2}g\delta\phi \gamma^T\sigma_y\gamma\right.\nonumber\\
&& \left.+\frac{1}{2}\delta\phi(-\partial^2+m^2)\delta\phi-\frac{g m}{2}\delta\phi^3+\frac{1}{8}g^2\delta\phi^4\right\}
\ .
}

At $r<0$ and $m=0$, $S^*$ reads
\eqn{
&& S^*=\int d\tau d^{d-1}x \left\{
\frac{1}{2}\gamma^T(i\partial_{\mu}\alpha^{\mu})\gamma+\frac{1}{2}g\phi \gamma^T\sigma_y\gamma\right.\nonumber\\
&& \left. +\frac{1}{2}\phi(-\partial^2+r)\phi+\frac{1}{8}g^2\phi^4\right\}
\ .
}
Since $r<0$, the boson part has spontaneous symmetry breaking(SSB) and the new vacua are at $\phi=\pm \sqrt{-2r}/g$.
Around the new vacua $\phi= \delta\phi\pm \sqrt{-2r}/g$, we have
\eqn{
\label{eq:S_m0}
&& S^*=\int d\tau d^{d-1}x \left\{
\frac{1}{2}\gamma^T(i\partial_{\mu}\alpha^{\mu}\pm \sqrt{-2r})\gamma+\frac{1}{2}g\delta\phi \gamma^T\sigma_y\gamma\right.\nonumber\\
&& \left. +\frac{1}{2}\delta\phi(-\partial^2-2 r)\delta\phi+\frac{\pm g \sqrt{-2 r}}{2}\delta\phi^3+\frac{1}{8}g^2\delta\phi^4\right\}
\ .
}

\section{Details on the Experimental Setup}
\label{app:ES}

In this section, we derive \eqnref{eq:S_setup} from \eqnref{eq:S_3D}.

To solve for the surface modes, we consider a semi-infinite configuration ($z<0$) with open boundary condition at $z=0$.
Then, we address the $z$ direction in the real space with
$\bar{\psi}_{\tau,\bsl{k}}=\int dz e^{i k_z z} \bar{\psi}_{\tau,\bsl{k}_\shpa,z}$, and \eqnref{eq:S_3D} becomes
\eqn{
&& S_0=\frac{1}{2\mathcal{S}_\shpa}\sum_{\bsl{k}_\shpa}\int d\tau \int_{-\infty}^0 d z\nonumber\\
&&\bar{\Psi}_{\tau,\bsl{k}_\shpa,z}(\partial_\tau
+h_{BdG}(\bsl{k}_\shpa,-i\partial_z))
\Psi_{\tau,\bsl{k}_\shpa,z}\ ,
}
where
\begin{equation}
h_{BdG}(\bsl{k}_\shpa,-i\partial_z)=
h_0(\bsl{k}_\shpa^2,-i\partial_z)+h_1(\bsl{k}_\shpa)\ ,
\end{equation}
\begin{equation}
h_0(\bsl{k}_\shpa^2,-i\partial_z)=\tau_z(-\frac{\partial_z^2}{2m_0}-\bar{\mu})+\Delta_p (-i\partial_z)\tau_x s_x\ ,
\end{equation}
\begin{equation}
h_1(\bsl{k}_\shpa)=\Delta_p (-\tau_x s_z) k_x+\Delta_p k_y (-\tau_y)\ ,
\end{equation}
$\bar{\mu}=\mu-\bsl{k}_\shpa^2/(2 m_0)$, $\tau_i$'s are Pauli matrices for the particle-hole index and $\bar{\Psi}_{\tau,\bsl{k}_\shpa,z}=(\bar{\psi}_{\tau,\bsl{k}_\shpa,z},\psi^{T}_{\tau,-\bsl{k}_\shpa,z})$.
In the following, we first solve for the zero modes of $h_0$ and then add $h_1$ as a perturbation since $\bsl{k}$ is small.
The zero mode equation for $h_0$ reads
\begin{equation}
\label{eq:zero_eq_Phi}
h_0(\bsl{k}_\shpa^2,-i\partial_z)\Phi_z=0\Leftrightarrow \left[-\bar{\mu}-\frac{\partial_z^2}{2m_0}+\Delta_p \tau_y s_x  \partial_z \right]\Phi_z=0\ .
\end{equation}
Define four orthonormal vectors $\xi_{a,j}$'s as
\begin{eqnarray}
&& \xi_{+,1}=(1,1,i,i)^T/2\nonumber\\
&& \xi_{+,2}=(1,-1,-i,i)^T/2\nonumber\\
&& \xi_{-,1}=(1,1,-i,-i)^T/2\nonumber\\
&& \xi_{-,2}=(1,-1,i,-i)^T/2\ ,
\end{eqnarray}
where $a=\pm$, $j=1,2$ and $\tau_y s_x \xi_{a,j}=a \xi_{a,j}$.
The wave function can be re-expressed as $\Phi_z=\sum_{a,j}f_{a,j}(z)\xi_{a,j}$, and \eqnref{eq:zero_eq_Phi} is equivalent to
\begin{equation}
(\partial_z^2-2 m_0 a \Delta_p \partial_z+2 m_0 \bar{\mu})f_{a,j}(z)=0
\end{equation}
for all $a=\pm$ and $j=1,2$ with boundary condition $f_{a,j}(0)=f_{a,j}(-\infty)=0$.
Without loss of generality, we choose $\Delta_p m_0 >0$.
In this case, we only have solution for $a=+$ and $2m_0\mu -\bsl{k}_\shpa^2>0$, which is $f_{+,j}(\bsl{k}_\shpa^2,z)=f_+(\bsl{k}_\shpa^2,z)=C_0 e^{m_0 \Delta_p z} \sinh(z\sqrt{\Delta_p^2 m_0^2-2 \bar{\mu} m_0})$ with $C_0$ the normalization constant that makes $f_{+}$ real.
The wave function of the zero modes in general has the form
\begin{equation}
\Phi_{\bsl{k}_\shpa^2,z}=f_{+}(\bsl{k}_\shpa^2,z)\sum_{j=1,2} C_j\xi_{+,j}\ .
\end{equation}
Clearly, there are two independent zero modes, of which the wavefunction can be chosen as
\begin{eqnarray}
&&\Phi_{1,\bsl{k}_\shpa^2,z}=\frac{f_+(\bsl{k}_\shpa^2,z)}{\sqrt{2}}(e^{-i\pi/4}\xi_{+,1}-e^{i\pi/4}\xi_{+,2})\nonumber\\
&&\Phi_{2,\bsl{k}_\shpa^2,z}=\frac{f_+(\bsl{k}_\shpa^2,z)}{\sqrt{2}}(-e^{-i\pi/4}\xi_{+,1}-e^{i\pi/4}\xi_{+,2})\ .
\end{eqnarray}
To get the low-energy effetive model, we define $\bar{\gamma}_{\tau,\bsl{k}_\shpa,i}=\int^0_{-\infty} d z \bar{\Psi}_{\bsl{k}_\shpa,z}\Phi_{i,\bsl{k}_\shpa^2,z}$.
Using  $\bar{\Psi}_{\tau,\bsl{k}_\shpa,z}=\sum_i \bar{\gamma}_{\tau,\bsl{k}_\shpa,i}\Phi^{\dagger}_{i,\bsl{k}_\shpa^2,z}+...$ with ``$...$'' the high energy contribution,
we can project $h_1$ to the zero modes and get
\begin{equation}
S_{eff}=\frac{1}{2 S_\shpa}\int d\tau\sum_{\bsl{k}_\shpa}\bar{\gamma}_{\tau,\bsl{k}_\shpa}[\partial_\tau+\Delta_p (k_x\sigma_z+k_y \sigma_x)]\gamma_{\tau,\bsl{k}_\shpa}\ ,
\end{equation}
where $\bar{\gamma}_{\tau,\bsl{k}_\shpa}=(\bar{\gamma}_{\tau,\bsl{k}_\shpa,1},\bar{\gamma}_{\tau,\bsl{k}_\shpa,2})$ and the terms related with the high-energy modes are neglected here.
Define $\gamma_{\tau,\bsl{k}_\shpa}=\int d^2 x e^{-i \bsl{k}_\shpa\cdot \bsl{x}}\gamma_{\tau,\bsl{x}}$.
Due to $\tau_x\Phi_{i,\bsl{k}_\shpa^2,z}^*=\Phi_{i,\bsl{k}_\shpa^2,z}$, $i\tau_0 s_y \Phi_{i,\bsl{k}_\shpa^2,z}^*=\sum_{j}\Phi_{j,\bsl{k}_\shpa^2,z}(i\sigma_y)_{ji}$, $\Psi^T_{\tau,-\bsl{k}_\shpa,z}=\bar{\Psi}_{\tau,\bsl{k}_\shpa,z}\tau_x$ and $\bar{\Psi}_{\tau,\bsl{k}_\shpa,z}\xrightarrow{\text{TR}}\Psi_{\tau,-\bsl{k}_\shpa,z}(i\tau_0 s_y)$, we have $\bar{\gamma}_{\tau,\bsl{x}}=\gamma_{\tau,\bsl{x}}^T$ and $\gamma_{\tau,\bsl{x}}^T\xrightarrow{\text{TR}} \gamma_{\tau,\bsl{x}}^T(i\sigma_y)$.
As a result, the above action becomes
\begin{equation}
S_{eff}=\frac{1}{2 }\int d\tau d^2 x \gamma_{\tau,\bsl{x}}^T[\partial_\tau+\Delta_p (-i\partial_x\sigma_z-i\partial_y \sigma_x)]\gamma_{\tau,\bsl{x}}\ .
\end{equation}
Replacing $\Delta_p$ by $v_f$ in the above action, we can get fermionic part of the \eqnref{eq:S_setup}.
Note that we neglect the high-energy modes, and thus $\Delta_p=v_f$ holds only to the leading order.

The surface magnetic doping of TSC can be phenomenologically described by the standard Ginzburg-Landau free energy of Ising magnetism, $S_M=\int d^3 x[\frac{1}{2} \phi(-\partial_\tau^2-v_b^2\bsl{\nabla}^2+r_0) \phi+\frac{1}{4!}u \phi^4]$, where $\phi_{\tau,x,y}$ is the order parameter of surface Ising magnetism along $z$.
$\phi$ is coupled to electrons at the surface through the exchange interaction $S_{ex}=\int d\tau \int d^3 r \phi_{\tau,x,y}g_M(z)  \bar{\psi}_{\tau,\bsl{r}}s_z \psi_{\tau,\bsl{r}}$ with $g_M(z)$ localized at the surface.
The matrix form of the exchange interaction in the BdG bases is $\tau_z s_z/2$, of which the projection to the Majorana modes is $\sigma_y/2$.
And thus the Ising coupling after the surface projection takes the form $\frac{1}{2}g\int d^3 x \phi \gamma^T \sigma_y \gamma$ with $g= \int dz f_+(0,z)^2 g_M(z)$.
Here we neglect the momentum dependence in $f_+(\bsl{k}_\shpa^2,z)$ as $\bsl{k}_\shpa$ is small.
Now the total action should read
\eqn{
&& S=\int d^{d}x \left[\frac{1}{2}\gamma^T(\partial_{\tau}-i v_f\bsl{\alpha}\cdot\bsl{\nabla})\gamma+\frac{1}{2}g \phi  \gamma^T\sigma_y\gamma\right.\nonumber\\
&&\left.+\frac{1}{2}\phi(-\partial_\tau^2-v_b^2\bsl{\nabla}^2+r_0)\phi+\frac{1}{4!}u \phi^4\right]\ ,
}
which has TR symmetry.
Furthermore, a magnetic field $B$ along $z$ is applied and coupled to both electron spin and Ising magnetism on the surface through the Zeeman-type action: $S_B=\int d\tau d^3 r g_s(z) B \bar{\psi}_{\tau,\bsl{r}} s_z \psi_{\tau,\bsl{r}} +\int d^3x ( a_1 B \phi+\frac{a_2 }{3!}B \phi^3)$ with $g_s(z)$ localized at the surface.
Here, we neglect the orbital effect as all fields are charge neutral, and add an $\phi^3$ term since it is allowed by symmetry and can be generated at the quantum level.
After projecting the $g_s$ term to the surface in the same way as the Ising coupling, we can include $S_B$ into the total action, and get \eqnref{eq:S_setup} with $\mu_B = \int dz f_+(0,z)^2  g_s(z)$.

%

\end{document}